\documentclass{aa}
\usepackage{txfonts, graphicx}
\usepackage[figuresright]{rotating}
\usepackage{longtable,lscape}
\usepackage{supertabular}
\usepackage{natbib}
\bibpunct{(}{)}{;}{a}{}{,}
\newcommand{\te}{\ensuremath{T_{\mathrm{eff}}}}
\newcommand{\kms}{km s$^{-1}$}

\newcommand{\lgg}{$\log g$}
\newcommand{\vsi}{\ensuremath{v \sin i}}
\newcommand{\bz}{\ensuremath{\langle B_z \rangle}}

\begin{document}
\titlerunning{Abundance variations in magnetic Ap stars}
\title{Discovery of secular variations in the atmospheric abundances of magnetic Ap stars\thanks{Based in part on observations made with the European Southern Observatory (ESO) telescopes under the ESO programmes 072.D-0410(A) and 086.D-0449(A). It is also based in part on observations carried out at the Canada-France-Hawaii Telescope (CFHT) which is operated by the National Research Council of Canada, the Institut National des Science de l\'Universe of the Centre National de la Recherche Scientifique of France and the University of Hawaii.}}
\author{J. D. Bailey\inst{1}
\and J. D. Landstreet\inst{2,1}
\and S. Bagnulo\inst{2}}
\institute{Department of Physics \& Astronomy, The University of Western Ontario, London, Ontario, N6A 3K7, Canada
\and
Armagh Observatory, College Hill, Armagh, Northern Ireland BT61 9DG}
\date{Accepted 26 November 2013}

\abstract{The stars of the middle main sequence have relatively quiescent outer layers, and unusual chemical abundance patterns may develop in their atmospheres. The presence of chemical peculiarities reveal the action of such subsurface phenomena as gravitational settling and radiatively driven levitation of trace elements, and their competition with mixing processes such as turbulent diffusion. At present, little is known about the time evolution of these anomalous abundances, nor about the role that diffusion may play in maintaining them, during the main sequence lifetime of such a star.}
{We want to establish whether abundance peculiarities change as stars evolve on the main sequence, and provide observational constraints to diffusion theory.}
{We have performed spectral analysis of 15 magnetic Bp stars that are members of open clusters (and thus have well-known ages), with masses between about 3 and 4~M$_{\odot}$. For each star, we measured the abundances of He, O, Mg, Si, Ti, Cr, Fe, Pr and Nd.}
{We have discovered the systematic time evolution of trace elements through the main-sequence lifetime of magnetic chemically peculiar stars as their atmospheres cool and evolve toward lower gravity. During the main sequence lifetime, we observe clear and systematic variations in the atmospheric abundances of He, Ti, Cr, Fe, Pr and Nd. For all these elements, except He, the atmospheric abundances decrease with age. The abundances of Fe-peak elements converge toward solar values, while the rare-earth elements converge toward values at least 100 times more abundant than in the Sun. Helium is always underabundant compared to the Sun, evolving from about 1\% up to 10\% of the solar He abundance. We have attempted to interpret the observed abundance variations in the context of radiatively driven diffusion theory, which appears to provide a framework to understand some, but not all, of the anomalous abundance levels and variations that we observe.}
{}

\maketitle

\section{Introduction}
Our empirical knowledge of the chemical history of the galaxy is based on the assumption that the atmospheres of the main sequence stars that trace its evolution have chemical compositions that reflect that of the interstellar gas at the time when they formed. This is not always the case, and it is essential to understand situations in which stellar surface chemistry does not reflect the bulk or initial chemical composition of the star. 

There are physical processes, such as diffusional gravitational settling of trace elements, and levitation of specific ions by radiative acceleration, that can lead to substantial evolution of atmospheric chemistry during the main sequence lifetime of a star. In cool stars (with atmospheric effective temperatures below about $T_{\rm eff} \le 7000$~K), these separation processes are overwhelmed by deep convective mixing of the outer envelope of the star, which maintains chemical homogeneity, and nearly the initial chemistry. In hot stars (with $T_{\rm eff} \ge 20\,000$~K), the intense outward radiation flux drives a strong stellar wind, stripping material off the stellar surface so rapidly that separation processes act too slowly to be able to compete.

Between these two effective temperature ranges lie the A--and late B--type main sequence stars. Such stars do not have any single powerful process acting to force the surface layers to retain essentially their initial chemical composition. In such stars, the atmospheric chemistry may evolve under the influence of loss of heavy trace ions downward by diffusive gravitational settling. Chemical evolution may also occur if the atmosphere acquires ions that are driven upwards from the invisible subsurface layers by radiative acceleration due to the outflowing radiative energy flux. Atoms that are driven into the atmosphere by this effect may also be driven out into space and thus lost \citep{Michaud1976}.

Other processes interact with or modify the effects of diffusion in the A and B stars. For example, a magnetic field can greatly impede the flow of material through the upper atmosphere of the star into interstellar space. Large-scale circulation currents in the stellar interior can lead to shear-induced turbulence that mixes subsurface layers. This mixing is strongly dependent on the stellar rotation rate, and is probably ineffective in stars with rotation periods of more than a few days. Accretion from dense interstellar clouds, through which the star passes from time to time, may also alter the surface chemistry \citep[e.g.][]{Havnes1971,Havnes1974,Kamp2002}. Finally, the star may be a member of a double system, and may acquire fresh surface material that is ejected by the evolution of its companion.

As a result, many A and B stars show chemical abundance ratios $\log N_{\rm X}/N_{\rm H}$ that are remarkably different from those measured in the Sun. Among A and B stars there are several families of distinctive compositional patterns that reflect different sets of physical conditions.  The compositional peculiarities vary rather strongly with effective temperature (and thus with the momentum carried by the outflowing radiation that levitates some trace elements). The peculiarities are very different, depending on whether the star has a strong magnetic field or not. There is also a family of peculiarities due to recent accretion from an interstellar cloud. These chemical peculiarities provide powerful probes of invisible processes occurring beneath the visible layers of all kinds of star, particularly of upward and downward diffusion \citep{Lan04}. It is therefore of obvious importance to observe and characterise the nature and time variations of these phenomena.

Until now, the study of atmospheric abundances of magnetic A and B (Ap and Bp) stars has been restricted mainly to individual studies of specific stars. The detail of these studies vary from coarse models that roughly describe the magnetic field geometry and abundance variations over the stellar surface \citep[e.g.][]{Landstreet1988,Bailey2011, Bailey2012} to more detailed maps of both the magnetic field structure and abundance distributions using spectroscopic observations in all four Stokes parameters \citep[e.g.][]{Koch2004,Koch2010}. In this paper, we present the results of a large project to study the time evolution of average surface characteristics in the Ap and Bp stars during the $10^8 - 10^9$~yrs of their main sequence phase. 

A major difficulty of determining how the observed chemical signatures vary with time has been that the ages of isolated (field) Ap and Bp
stars can be estimated only very roughly. To solve this problem we have studied surface abundance evolution in a sample of magnetic peculiar A and B stars that are open cluster members. The ages of cluster members can be determined with much better accuracy than those of isolated stars, and the cluster age applies to all its members, which are presumed to have formed essentially contemporaneously \citep{Meyetal93,paper1,paper2}. 

The following section discusses the observations. Sect.~3 describes the modelling technique. Sect.~4 details the observational results. Sect.~5 explores possible mechanisms to explain the observed trends and Sect.~6 summarises the main conclusions.  

\section{Observations}
\begin{table}
\centering
\caption{Stars analysed in this study.  Listed are the star designations, instrument used (with the number of spectra indicated in parentheses), spectral resolution, and spectral range.}
\begin{tabular}{lrrr}
\hline\hline
Star & Instrument & R & $\lambda$ (\AA) \\
\hline
HD 45583 & FEROS (2) & 48000 & 3528-9217 \\
HD 61045 & ESPaDOnS (2) & 65000 & 3690-10481 \\ 
HD~63401  & FEROS (2) & 48000 & 3528-9217 \\
HD 74535 & FEROS (2) & 48000 & 3528-9217 \\  
HD~133652 & ESPaDOnS (1) & 65000 & 3690-10481 \\
 & FEROS (1) & 48000 & 3528-9217 \\ 
HD 133880 & ESPaDOnS (2) & 65000 & 3690-10481 \\ 
HD 147010 & ESPaDOnS (2) & 65000 & 3690-10481 \\
HD~162576 & ESPaDOnS (2) & 65000 & 3690-10481 \\
HD~162725 & ESPaDOnS (2) & 65000 & 3690-10481 \\
HD 304842 & FEROS (2) & 48000 & 3528-9217 \\
NGC~2169~12 & UVES (1) & 110000 & 3070-10398 \\
BD+00 1659 & ESPaDOnS (1) & 65000 & 3690-10481 \\
BD-19 5044L & ESPaDOnS (2) & 65000 & 3690-10481 \\
BD+49 3789 & ESPaDOnS (2) & 65000 & 3690-10481 \\ 
HIP~109911 & ESPaDOnS (2) & 65000 & 3690-10481 \\
\hline
\label{observations}
\end{tabular}
\end{table}
Our study used high dispersion spectra of 15 stars, listed in Table~\ref{observations}. The majority of spectra were acquired using the ESPaDOnS spectropolarimeter and the FEROS spectrograph located at the Canada-France-Hawaii-Telescope (CFHT) and the European Southern Observatory's (ESO) La Silla Observatory, respectively. One spectrum was acquired using the UVES spectrograph at ESO's Paranal Observatory.

We are studying a sample of stars with a limited range of masses (between 3 and 4~M$_{\odot}$) because the physical effects driving atmospheric abundance changes, and the main sequence time-scale over which these effects can act, are expected to depend on mass. Our sample is selected to allow us to reconstruct the atmospheric abundance evolution of a magnetic Bp star of about $3.5 M_\odot$. 

The stars of our sample are all definite or probable members of open clusters or associations. The age of each cluster was determined by fitting a theoretical isochrone to the cluster members in an HR diagram ($\log (L/L_{\odot})$ versus $T_{\rm eff}$). For each cluster, the brightest members (those closest to turn-off from the main sequence) are used as the main indicators of cluster age. It has been found that fitting isochrones to the observed stars in the theoretical HR diagram provides a more precise cluster age than doing this fitting in an observational HR diagram, such as $V$ versus $B-V$. This is because in the theoretical HR diagram the isochrones have a strong hook near the cluster turn-off, while in the observational HR diagram the isochrones tend to a vertical line that does not discriminate clearly between various ages \citep{paper2}.

For each star, accurate values of effective temperature ($T_{\rm eff}$), gravity ($\log g$), evolutionary mass, magnetic field strength and age were adopted from the literature
\citep{paper2,BL2013}. When values of $T_{\rm eff}$ and $\log g$ were not available, we derived them from available Geneva and $uvby\beta$ photometry. For the Geneva photometry the {\sc fortran} program described by \citet{geneva} was used. For the Str\"{o}mgren $uvby\beta$ photometry we used a version of the {\sc fortran} program ``UVBYBETANEW'' \citep[see][]{NSW} that corrects the \te\ of the magnetic Bp stars to the appropriate temperature scale \citep[see][]{paper2}. As per the discussions of \citet{paper2} and \citet{BL2013}, the uncertainties in \te\ and \lgg\ were taken to be about $\pm$500~K and 0.2~dex, respectively. Table~\ref{properties} lists the properties of each star, including its designation, associated cluster, age, mass, $T_{\rm eff}$, $\log g$, and $v \sin i$. The final column shows the root-mean-square magnetic field strength ($B_{\rm rms}$) as computed from whatever modern measurements of mean longitudinal magnetic field $B_{\rm z}$ are available \citep{paper2,paper3}; this quantity is the most useful value we have to characterise the magnetic field of each star. Where applicable, the appropriate reference is given for each parameter. Unreferenced parameters were derived for this study. We note that the photometrically determined values of $T_{\rm eff}$ and $\log g$ vary systematically with increasing age from about 13500~K and 4.4 to 10000~K and 3.5, as expected for the evolution of a single star of about 3.5~M$_{\odot}$ from ZAMS to TAMS.

\section{Modelling technique}
\begin{center}
\begin{table*}
\caption{Physical properties of the stars studied.}
\begin{tabular}{lrlrrrrr}
\hline\hline
Star & Cluster & $\log t$ & M/M$_{\odot}$ & \te\ (K) & \lgg\ & \vsi\ (\kms) & B$_{\rm rms}$ (G)\\  
\hline
HD~147010 & Upper Sco & 6.70 $\pm$ 0.10$^{1}$ & 3.15 $\pm$ 0.20$^{1}$ & 13000 $\pm$ 500$^{2}$ & 4.40 $\pm$ 0.20$^{2}$ & 15 $\pm$ 2$^{2}$ & 4825$^{1}$ \\
NGC~2169~12 & NGC~2169 & 6.97 $\pm$ 0.10$^{1}$ & 3.65 $\pm$ 0.15$^{1}$ & 13800 $\pm$ 500$^{1}$ & 4.30 $\pm$ 0.20 & 56 $\pm$ 5 & 3410$^{1}$\\
HD~133652 & Upper Cen Lup & 7.20 $\pm$ 0.10$^{1}$ & 3.35 $\pm$ 0.15$^{1}$ & 13000 $\pm$ 500 & 4.30 $\pm$ 0.20 & 48 $\pm$ 2 & 1120$^{1}$\\
HD~133880 & & & 3.20 $\pm$ 0.15$^{1}$ & 13000 $\pm$ 600$^{3}$ & 4.34 $\pm$ 0.16$^{3}$ & 103 $\pm$ 10$^{3}$ & 2300$^{1}$ \\
HD~45583 & NGC~2232 & 7.55 $\pm$ 0.10$^{1}$ & 3.30 $\pm$ 0.15$^{1}$ & 12700 $\pm$ 500$^{2}$ & 4.20 $\pm$ 0.20$^{2}$ & 70 $\pm$ 6$^{2}$ & 2730$^{1}$\\
HD~63401 & NGC~2451 & 7.70 $\pm$ 0.10$^{1}$ & 3.70 $\pm$ 0.20$^{1}$ & 13500 $\pm$ 500$^{1}$ & 4.20 $\pm$ 0.20 & 52 $\pm$ 4 & 365$^{1}$ \\
HD~74535 & IC~2391 & 7.70 $\pm$ 0.15$^{1}$ & 3.85 $\pm$ 0.15$^{1}$ & 13600 $\pm$ 500$^{1}$ & 4.30 $\pm$ 0.20 & 45 $\pm$ 4 & 95$^{1}$\\
BD-19~5044L & IC~4725 & 8.02 $\pm$ 0.08$^{1}$ & 3.55 $\pm$ 0.15$^{1}$ & 12800 $\pm$ 500$^{2}$ & 4.50 $\pm$ 0.20$^{2}$ & 15 $\pm$ 3$^{2}$ & 235$^{1}$\\
BD+49~3789 & NGC~7243 & 8.06 $\pm$ 0.10$^{*}$ & 3.55 $\pm$ 0.15$^{*}$ & 12900 $\pm$ 500$^{2}$ & 4.20 $\pm$ 0.20$^{2}$ & 85 $\pm$ 5$^{2}$ & 561$^{*}$\\
HIP~109911 & & & 3.65 $\pm$ 0.15$^{*}$ & 13000 $\pm$ 500 & 4.30 $\pm$ 0.20 & 60 $\pm$ 2 & 348$^{*}$\\  
HD~61045 & NGC~2422 & 8.08 $\pm$ 0.11$^{1}$ & 3.85 $\pm$ 0.20$^{1}$ & 13000 $\pm$ 500$^{2}$ & 4.10 $\pm$ 0.20$^{2}$ & 64 $\pm$ 3$^{2}$ & 430$^{1}$\\
HD~304842 & NGC~3114 & 8.13 $\pm$ 0.15$^{1}$ & 3.55 $\pm$ 0.15$^{1}$ & 12500 $\pm$ 500$^{2}$ & 3.90 $\pm$ 0.20$^{2}$ & 65 $\pm$ 5$^{2}$ & 20$^{1}$\\
BD+00~1659 & NGC~2301 & 8.22 $\pm$ 0.10$^{*}$ & 3.65 $\pm$ 0.15$^{*}$ & 12500 $\pm$ 500$^{2}$ & 4.00 $\pm$ 0.20$^{2}$ & 7.0 $\pm$ 1$^{2}$ & 394$^{*}$\\
HD~162576 & NGC~6475 & 8.41 $\pm$ 0.13$^{1}$ & 3.10 $\pm$ 0.15$^{*}$ & 10300 $\pm$ 500 & 3.70 $\pm$ 0.20 & 28 $\pm$ 3 & 15$^{*}$\\
HD~162725 & & & 3.30 $\pm$ 0.20$^{1}$ & 10000 $\pm$ 500 & 3.50 $\pm$ 0.20 & 31 $\pm$ 3 & 69$^{*}$\\
\hline
\multicolumn{8}{p{0.85\textwidth}}{{\sc references} -- (1) \citet{paper2}; (2) \citet{BL2013}; (3) \citet{Bailey2012}; (*) This work} \\
\label{properties}
\end{tabular}
\end{table*}
\end{center}  

To determine the atmospheric abundances of the magnetic Bp stars, the {\sc fortran} program {\sc zeeman} was used \citep{Landstreet1988,Wadeetal2001}. {\sc zeeman} is a spectrum synthesis program for stars having magnetic fields. It assumes a magnetic field geometry which is modelled as a simple co-linear multipole expansion, with the strength of the multipole components, the inclination $i$ of the rotation axis, and the obliquity $\beta$ of the magnetic field axis to the rotation axis specified. {\sc zeeman} interpolates an appropriate stellar atmospheric structure from a pre-tabulated grid of ATLAS 9 atmospheric models, based on the $T_{\rm eff}$ and $\log g$ assumed. The atomic data for individual spectral lines are taken from the Vienna Atomic Line Database \citep[VALD;][]{vald1, vald2, vald3, vald4}. For all stars, a uniform atmospheric abundance distribution vertically and over the stellar surface was assumed. The microturbulence parameter has been set to zero for all abundance determinations for two reasons. Firstly, it has been found that even non-magnetic stars show no evidence of microturbulence between about 11000 and 14000~K \citep{Lanetal09}; and secondly, it is very probable that the magnetic field is able to suppress convective motions in the atmosphere because of the large energy density in the field.

{\sc zeeman} searches for an optimal fit between the synthetic and observed spectra by means of a reduced $\chi^{2}$ fit of the computed spectrum to the observed one. Multiple spectral windows can be synthesised simultaneously and {\sc zeeman} automatically provides as output the best values for the radial velocity $v_{\rm R}$ and $v \sin i$. The abundance ($\log N_{\rm X}/N_{\rm H}$) of one element at a time is optimised by identifying unblended lines of that element and then fitting, as well as possible, the observed spectral lines of that element. The stars modelled vary in $T_{\rm eff}$ from about 10000 to 14000~K and therefore share many spectral lines in common. For consistency, we endeavoured to deduce elemental abundances from the same sets of lines for all stars. Table~\ref{lines} lists the lines used for modelling all the stars. As far as possible, lines with a range of strengths were used to deduce the final abundances. For each element, we adopted an uncertainty consistent with the change in abundance from the best-fit model that was necessary to produce an unsatisfactory fit in the spectral window (determined by visual inspection). {\sc zeeman} cannot model simultaneously spectral lines that are widely separated. In such instances, at least two lines (in different spectral windows) were fit separately and the average abundance deduced from the lines was adopted. The uncertainty in this case was estimated from the observed scatter between the computed values of the different spectral lines.
\begin{table}
\caption{List of spectral lines modelled.}
\begin{tabular}{lrlr}
\hline\hline
Element & $\lambda$ (\AA) & Element & $\lambda$ (\AA) \\  
\hline
He~{\sc i} & 4437.551 & Si~{\sc ii} & 5055.984 \\
& 4713.139 & & 5056.317\\
& 5015.678 & Ti~{\sc ii} & 4533.960\\
& 5047.738 & & 4563.757 \\
& 5875.599 & & 4571.968 \\
& 5875.614 & Cr~{\sc ii} & 4558.650  \\
& 5875.615 & & 4565.739  \\
& 5875.625 & & 4588.199 \\
& 5875.640 & & 4592.049 \\
& 5875.966 & Fe~{\sc ii} & 4541.524\\ 
O~{\sc i} & 6155.961 & & 4555.893\\ 
& 6155.966 & & 4556.392 \\
& 6155.986 & & 4583.837 \\
& 6156.736 & & 4583.999\\
& 6156.755 & & 5029.097 \\
& 6156.776 & & 5030.630 \\
& 6158.146 & & 5032.712 \\
& 6158.172 & & 5035.708 \\
& 7771.941 &  Pr~{\sc iii} & 6160.233 \\
& 7774.161 & & 6161.194 \\
& 7775.388 & & 7781.983 \\
Mg~{\sc ii} & 4481.126 & Nd~{\sc iii} & 4911.653 \\ 
& 4481.150 & & 4912.944 \\ 
& 4481.325 & & 4914.094 \\
Si~{\sc ii} & 4621.418 & & 5050.695 \\ 
& 4621.722 & & 6145.068 \\
& 5041.024 & & \\
\hline
\label{lines}
\end{tabular}
\end{table}
\begin{center}
\begin{figure*}
\centering
\includegraphics[angle=-90, width=0.95\textwidth]{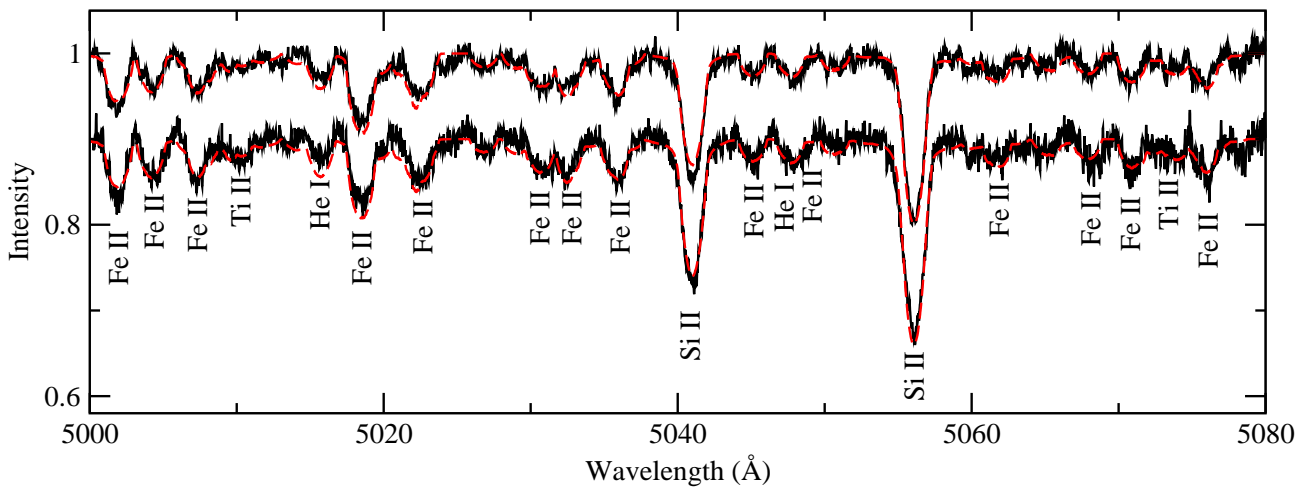}
\caption{Spectrum synthesis of the region 5000 - 5080~\AA\ for two spectra of HIP~109911. The observed spectra are in black and the model fits are in red.
}
\label{sample}
\end{figure*}
\end{center}
All of our stars have measurements of the line-of-sight magnetic field \bz, either previously published \citep{paper1,Kochukhov2006,paper3} or from our own unpublished
results. For two of the stars in our sample, magnetic field models are available from detailed studies: HD~133880 \citep{Bailey2012} and HD~147010 \citep{Bailey2013}. In these cases, we adopted the published geometries. When no detailed magnetic field model was available, we simply assumed a dipolar magnetic field that was approximately three
times $B_{\rm rms}$ and computed the spectrum with the line of sight parallel to the magnetic axis. For the majority of the stars in our sample we have multiple observations. In the cases where more than two spectra were available, we chose the two that exhibited the greatest differences. In all cases, the derived abundances are the average of results from the two modelled spectra, with the associated uncertainties propagated accordingly.

\subsection{Abundance analysis}
Atmospheric abundances of He, O, Mg, Si, Ti, Cr, Fe, Pr and Nd were determined for the 15 cluster magnetic Bp stars in our 3.5~M$_{\odot}$ sample. In Table~\ref{abundances}, we tabulate the mean abundances {\bf $\log (N_{\rm X}/N_{\rm H})$} for each star. The first three columns recall the star designation, age ($\log t$) and $T_{\rm eff}$ from Table~\ref{properties}. The subsequent columns list the average abundance values, with their associated uncertainties, for He, O, Mg, Si, Ti, Cr, Fe, Pr and Nd. For reference, the solar abundance ratios are also shown \citep{Asplund2009}. Note that the abundance scale we use can be converted to the common scale, having the logarithmic abundance of H as +12, by adding 12 to all our values. Figure~\ref{sample} provides an example of the quality of fits we achieved for each star.

We know that serious discrepancies (of the order of 1~dex) are found between abundances derived from lines of Si~{\sc ii} compared to those derived using lines of Si~{\sc iii}, and that this situation exists in most, or all, magnetic stars within the $T_{\rm eff}$ range of this study \citep{BL2013}. In general, for magnetic Bp stars the abundance of Si deduced from lines of Si~{\sc iii} is significantly larger than the value found using lines of Si~{\sc ii}. The study by \citet{BL2013} suggests that this discrepancy is mostly due to strong vertical stratification of Si, similar to the stratification profiles computed by \citep{LeBlanc2009} for stellar atmospheres with $T_{\rm eff} = 8000$ and 12000~K, and shown in their Figure 9. We have determined Si abundances using only lines of Si~{\sc ii}.

Stratification has been directly deduced from observed spectra in the vertical distribution of Fe and other elements \citep[e.g.,][]{Bagetal01,Wade2001b,Ryab2005b}, mainly in cool magnetic Ap stars. We expect this phenomenon to occur in the $T_{\rm eff}$ range discussed in this paper as well. Because both observation \citep{Bagetal01,Ryab2004b} and theory \cite{LeBlanc2009} suggest that the vertical abundance variations may be of the order of 1~dex or more, this phenomenon raises quite serious questions about the precise meaning of atmospheric chemical abundances derived assuming that the elements are uniformly distributed in the vertical direction. This is similar to the ambiguous meaning of abundances derived for (horizontally) patchy stars using models that assume uniform horizontal abundance.

Although these complications make the meaning of abundance determinations rather inexact, such simplified abundance measurements nevertheless form a useful ``instrumental'' system, which reveals differences between different stars satisfactorily. The typical symptom of vertical abundance stratification is that when we find a homogeneous model spectrum that fits lines of an element of intermediate strength, that same model predicts lines that are stronger than the strongest observed lines, and weaker than the weakest observed lines (see Figs.~2 of \citealt{Bagetal01}). Thus, by simply fitting on average the spectral lines of the element under study, we hypothesise that we are deriving a fairly stable mean abundance which has a similar meaning in stars that do not differ greatly in \te. However, the ambiguity due to probable vertical stratification and horizontal inhomogeneities should be kept in mind in evaluating our results.

\begin{sidewaystable*}
\vspace{20cm}
\caption{Average derived abundances for the stars studied.}
\centering
\begin{tabular}{llllllllllll}
\hline\hline
Star & $\log t$ & \te\ (K) & $\log$(He/H) & $\log$(O/H) & $\log$(Mg/H) & $\log$(Si~{\sc ii}/H) & $\log$(Ti/H) & $\log$(Cr/H) & $\log$(Fe/H) & $\log$(Pr/H) & $\log$(Nd/H) \\  
\hline
HD~147010 & 6.70 $\pm$ 0.10$^{1}$ & 13000 $\pm$ 500$^{2}$ & $-3.11 \pm 0.28$ & $-4.49 \pm 0.28$ & $-5.67 \pm 0.14$ & $-3.72 \pm 0.14$ & $-5.68 \pm 0.21$ & $-4.04 \pm 0.14$ & $-3.33 \pm 0.14$ & $-6.15 \pm 0.28$ & $-6.50 \pm 0.28$ \\
NGC~2169~12 & 6.97 $\pm$ 0.10$^{1}$ & 13800 $\pm$ 500$^{1}$ & $-3.09 \pm 0.30$ & $-3.88 \pm 0.15$ & $-5.56 \pm 0.15$ & $-3.87 \pm 0.15$ & $-5.90 \pm 0.20$ & $-4.80 \pm 0.20$ & $-3.76 \pm 0.10$ & $-7.31 \pm 0.20$ & $-7.21 \pm 0.20$ \\
HD~133652 & 7.20 $\pm$ 0.10$^{1}$ & 13000 $\pm$ 500 & $-3.16 \pm 0.28$ & $-3.87 \pm 0.28$ & $-5.18 \pm 0.14$ & $-3.36 \pm 0.28$ & $-5.01 \pm 0.21$ & $-4.17 \pm 0.21$ & $-2.90 \pm 0.14$ & $-6.98 \pm 0.28$ & $-6.42 \pm 0.28$ \\
HD~133880 & & 13000 $\pm$ 600$^{3}$ & $\leq-2.00$ & $-2.90 \pm 0.42$ & $-4.12 \pm 0.28$ & $-2.94 \pm 0.28$ & $-5.44 \pm 0.28$ & $-4.65 \pm 0.21$ & $-3.56 \pm 0.14$ & $-6.96 \pm 0.28$ & $-7.04 \pm 0.42$ \\
HD~45583 & 7.55 $\pm$ 0.10$^{1}$ & 12700 $\pm$ 500$^{2}$ & $\leq-2.40$ & $-3.61 \pm 0.42$ & $-4.19 \pm 0.28$ & $-3.34 \pm 0.28$ & $-5.68 \pm 0.28$ & $-4.74 \pm 0.21$  & $-3.43 \pm 0.21$ & $-7.49 \pm 0.42$ & $-7.04 \pm 0.42$ \\
HD~63401 & 7.70 $\pm$ 0.10$^{1}$ & 13500 $\pm$ 500$^{1}$ & $-2.64 \pm 0.28$ & $-3.73 \pm 0.21$ & $-5.86 \pm 0.14$ & $-3.96 \pm 0.14$ & $-6.44 \pm 0.28$ & $-5.32 \pm 0.28$ & $-3.84 \pm 0.14$ & $-7.14 \pm 0.28$ & $-7.44 \pm 0.28$ \\
HD~74535 & 7.70 $\pm$ 0.15$^{1}$ & 13600 $\pm$ 500$^{1}$ & $-2.52 \pm 0.28$ & $-4.26 \pm 0.28$ & $-5.44 \pm 0.28$ & $-4.28 \pm 0.14$ & $-6.29 \pm 0.28$ & $-5.65 \pm 0.14$ & $-4.02 \pm 0.28$ & $-8.15 \pm 0.21$ & $-7.52 \pm 0.28$ \\
BD-19~5044L & 8.02 $\pm$ 0.08$^{1}$ & 12800 $\pm$ 500$^{2}$ & $-2.16 \pm 0.21$ & $-3.99 \pm 0.28$ & $-5.44 \pm 0.21$ & $-3.94 \pm 0.28$ & $-6.82 \pm 0.28$ & $-5.88 \pm 0.28$ & $-4.24 \pm 0.28$ & $-8.12 \pm 0.28$ & $-7.88 \pm 0.42$ \\
BD+49~3789 & 8.06 $\pm$ 0.10$^{*}$ & 12900 $\pm$ 500$^{2}$ & $-2.52 \pm 0.35$ & $-3.60 \pm 0.22$ & $-5.25 \pm 0.25$ & $-3.54 \pm 0.14$ & $-6.33 \pm 0.35$ & $-5.44 \pm 0.28$ & $-4.06 \pm 0.28$ & $-7.72 \pm 0.28$ & $-7.72 \pm 0.25$ \\
HIP~109911 & & 13000 $\pm$ 600 & $-2.34 \pm 0.28$ & $-3.69 \pm 0.21$ & $-5.38 \pm 0.14$ & $-3.49 \pm 0.14$ & $-5.90 \pm 0.28$ & $-5.35 \pm 0.28$ & $-3.59 \pm 0.28$ & $-7.70 \pm 0.28$ & $-7.37 \pm 0.28$ \\  
HD~61045 & 8.08 $\pm$ 0.11$^{1}$ & 13000 $\pm$ 500$^{2}$ & $-2.06 \pm 0.28$ & $-3.72 \pm 0.28$ & $-5.10 \pm 0.14$ & $-3.94 \pm 0.28$ & $-6.24 \pm 0.28$ & $-5.47 \pm 0.28$ & $-3.98 \pm 0.28$ & $-7.74 \pm 0.28$ & $-7.77 \pm 0.42$ \\
HD~304842 & 8.13 $\pm$ 0.15$^{1}$ & 12500 $\pm$ 500$^{2}$ & $-1.54 \pm 0.28$ & $-3.78 \pm 0.25$ & $-5.76 \pm 0.21$ & $-3.64 \pm 0.28$ & $-6.83 \pm 0.14$ & $-6.05 \pm 0.32$ & $-4.35 \pm 0.32$ & $-7.65 \pm 0.36$ & $-7.04 \pm 0.28$ \\
BD+00~1659 & 8.22 $\pm$ 0.10$^{*}$ & 12500 $\pm$ 500$^{2}$ & $-2.34 \pm 0.15$ & $-3.58 \pm 0.10$ & $-5.62 \pm 0.10$ & $-3.53 \pm 0.20$ & $-6.69 \pm 0.10$ & $-5.02 \pm 0.10$ & $-3.79 \pm 0.10$ & $-8.11 \pm 0.20$ & $-7.34 \pm 0.20$ \\
HD~162576 & 8.41 $\pm$ 0.13$^{1}$ & 10300 $\pm$ 500 & $-1.51 \pm 0.28$ & $-3.77 \pm 0.28$ & $-5.34 \pm 0.14$ & $-4.44 \pm 0.14$ & $-7.72 \pm 0.14$ & $-5.12 \pm 0.14$ & $-3.95 \pm 0.14$  & $-10.15 \pm 0.42$ & $-9.10 \pm 0.42$ \\
HD~162725 & & 10000 $\pm$ 500 & $-1.74 \pm 0.36$ & $-3.77 \pm 0.28$ & $-5.23 \pm 0.14$ & $-3.54 \pm 0.28$  & $-7.30 \pm 0.28$ & $-4.64 \pm 0.22$ & $-3.66 \pm 0.28$ & $-8.32 \pm 0.42$ & $-7.44 \pm 0.28$ \\
Sun & & & $-1.07$ & $-3.31$ & $-4.40$ & $-4.49$ & $-7.05$ & $-6.36$ & $-4.50$ & $-11.28$ & $-10.58$\\
\hline
\multicolumn{12}{p{0.95\textwidth}}{{\sc references} -- (1) \citet{paper2}; (2) \citet{BL2013}; (3) \citet{Bailey2012}; (*) This work}\\
\label{abundances}
\end{tabular}
\end{sidewaystable*}

\section{Results}
\begin{center}
\centering
\begin{table}
\caption{The linear fit parameters for the abundance of each element versus age and magnetic field strength. Shown are the slopes with their respective uncertainties and the significance of the slope, $\sigma_{D}$  }
\begin{tabular}{llrr}
\hline\hline
y-axis & x-axis & Slope & $\sigma_{D}$\\ 
\hline
 $\log(\rm He/H)$ & $\log t$ & 0.76 $\pm$ 0.18 & 4.2\\ 
 & $\log B_{\rm rms}$ & $-0.48$ $\pm$ 0.11 & 4.4 \\ 
 $\log(\rm O/H)$ & $\log t$ & 0.13 $\pm$ 0.17 & 0.76\\
 & $\log B_{\rm rms}$ & 0.04 $\pm$ 0.13 & 0.31\\ 
 $\log(\rm Mg/H)$ & $\log t$ & $-0.15$ $\pm$ 0.26 & 0.58\\ 
 & $\log B_{\rm rms}$ & 0.20 $\pm$ 0.18 & 1.1\\ 
 $\log(\rm Si/H)$ & $\log t$ & $-0.13$ $\pm$ 0.20 & 0.65\\ 
& $\log B_{\rm rms}$ & 0.19 $\pm$ 0.13 & 1.5\\ 
 $\log(\rm Ti/H)$ & $\log t$ & $-1.07$ $\pm$ 0.27 & 4.0\\ 
& $\log B_{\rm rms}$ & 0.69 $\pm$ 0.12 & 5.8\\ 
 $\log(\rm Cr/H)$ & $\log t$ & $-0.97$ $\pm$ 0.24 & 4.0\\ 
& $\log B_{\rm rms}$ & 0.40 $\pm$ 0.16 & 2.5\\ 
 $\log(\rm Fe/H)$ & $\log t$ & $-0.41$ $\pm$ 0.16 & 2.6\\ 
& $\log B_{\rm rms}$ & 0.24 $\pm$ 0.08 & 3.0 \\ 
 $\log(\rm Pr/H)$ & $\log t$ & $-1.29$ $\pm$ 0.29 & 4.4\\ 
& $\log B_{\rm rms}$ & 0.88 $\pm$ 0.20 & 4.4\\
 $\log(\rm Nd/H)$ & $\log t$ & $-0.81$ $\pm$ 0.24 & 3.4\\ 
& $\log B_{\rm rms}$ & 0.46 $\pm$ 0.17 & 2.7\\ 
\hline
\label{lin-fits}
\end{tabular}
\end{table}
\end{center}  

\subsection{Abundance variations with time}
\begin{figure*}
\centering
\includegraphics[angle=-90,width=0.33\textwidth]{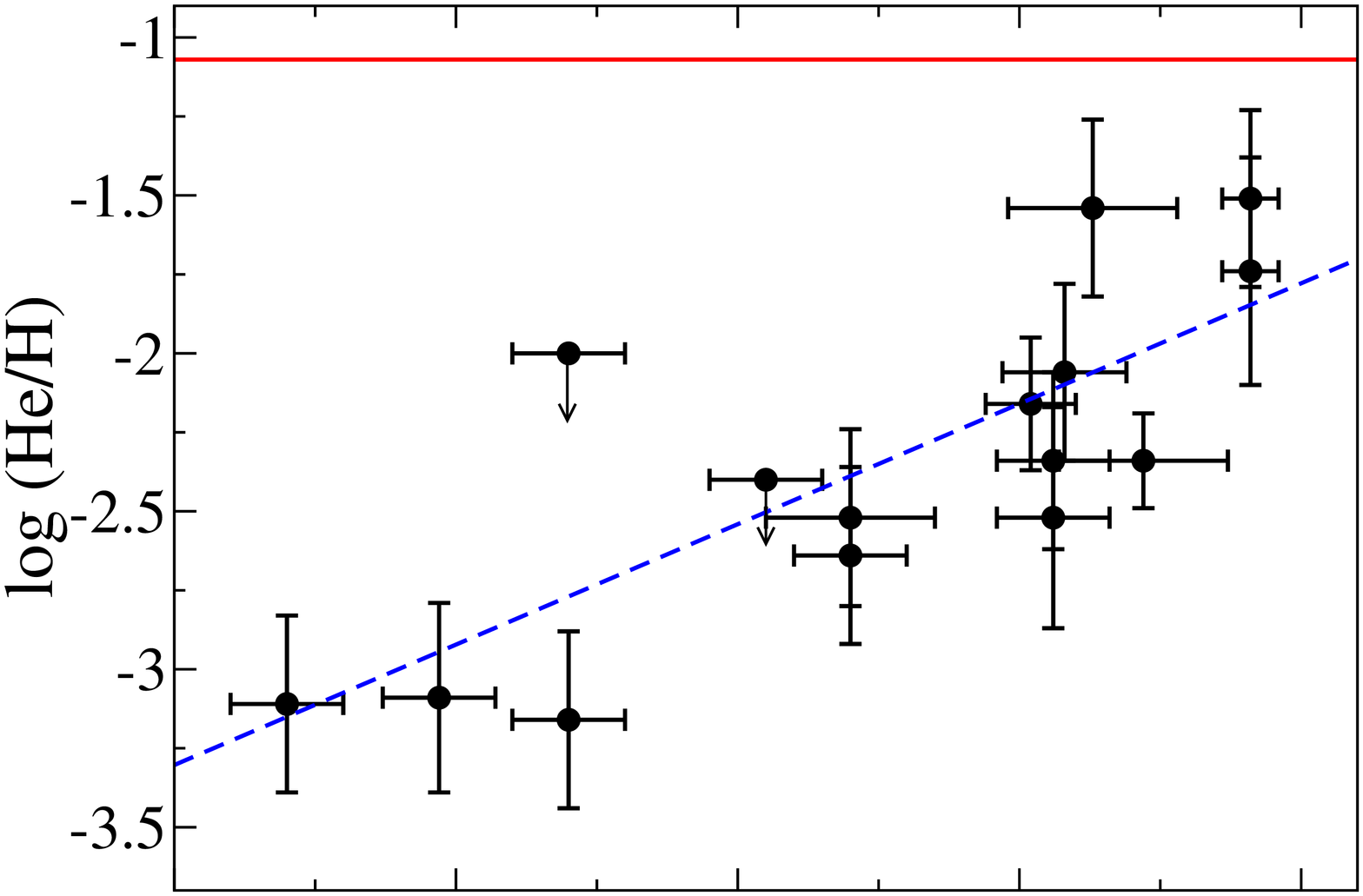}
\includegraphics[angle=-90,width=0.33\textwidth]{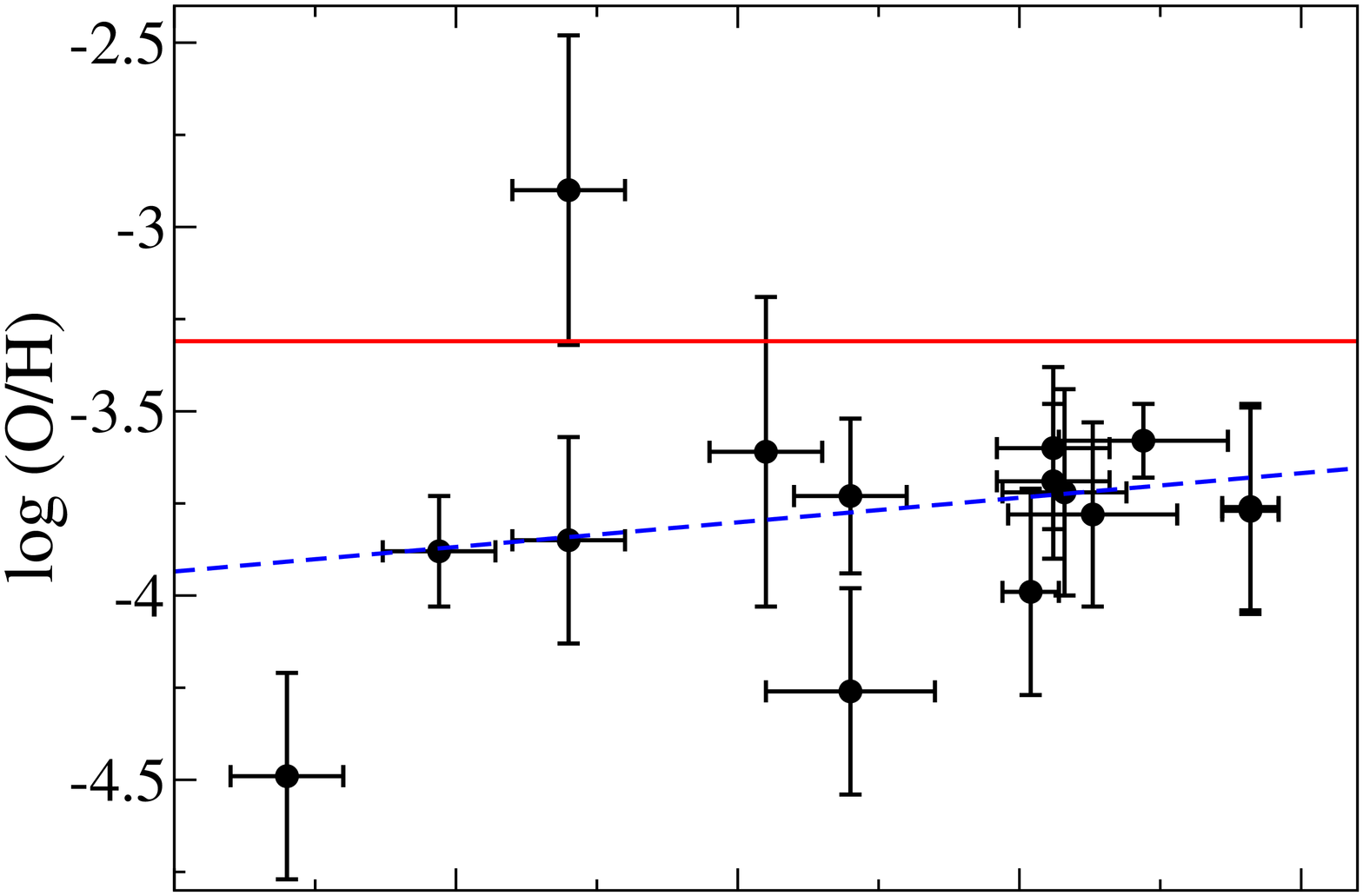}
\includegraphics[angle=-90,width=0.33\textwidth]{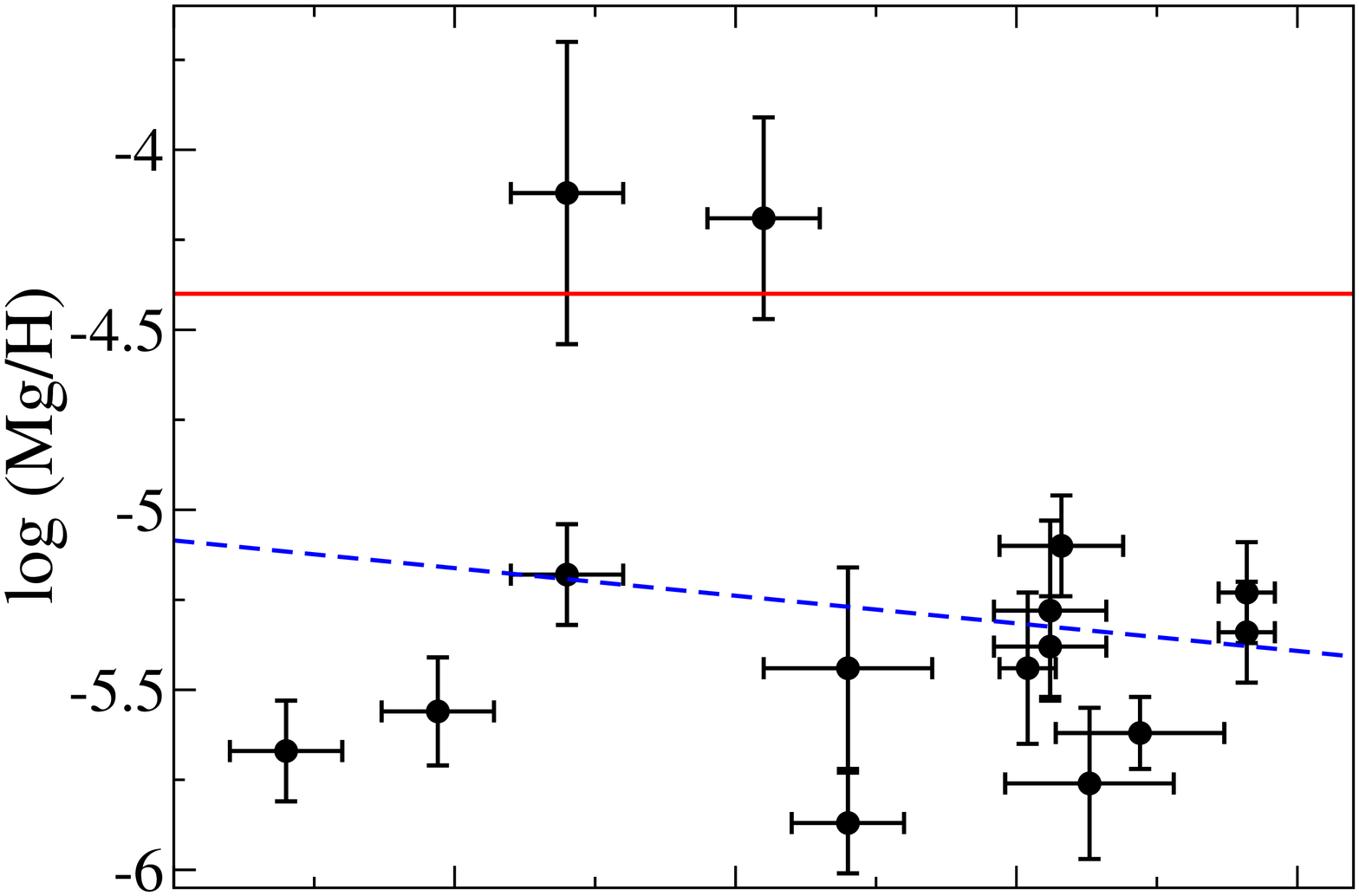}\\
\includegraphics[angle=-90,width=0.33\textwidth]{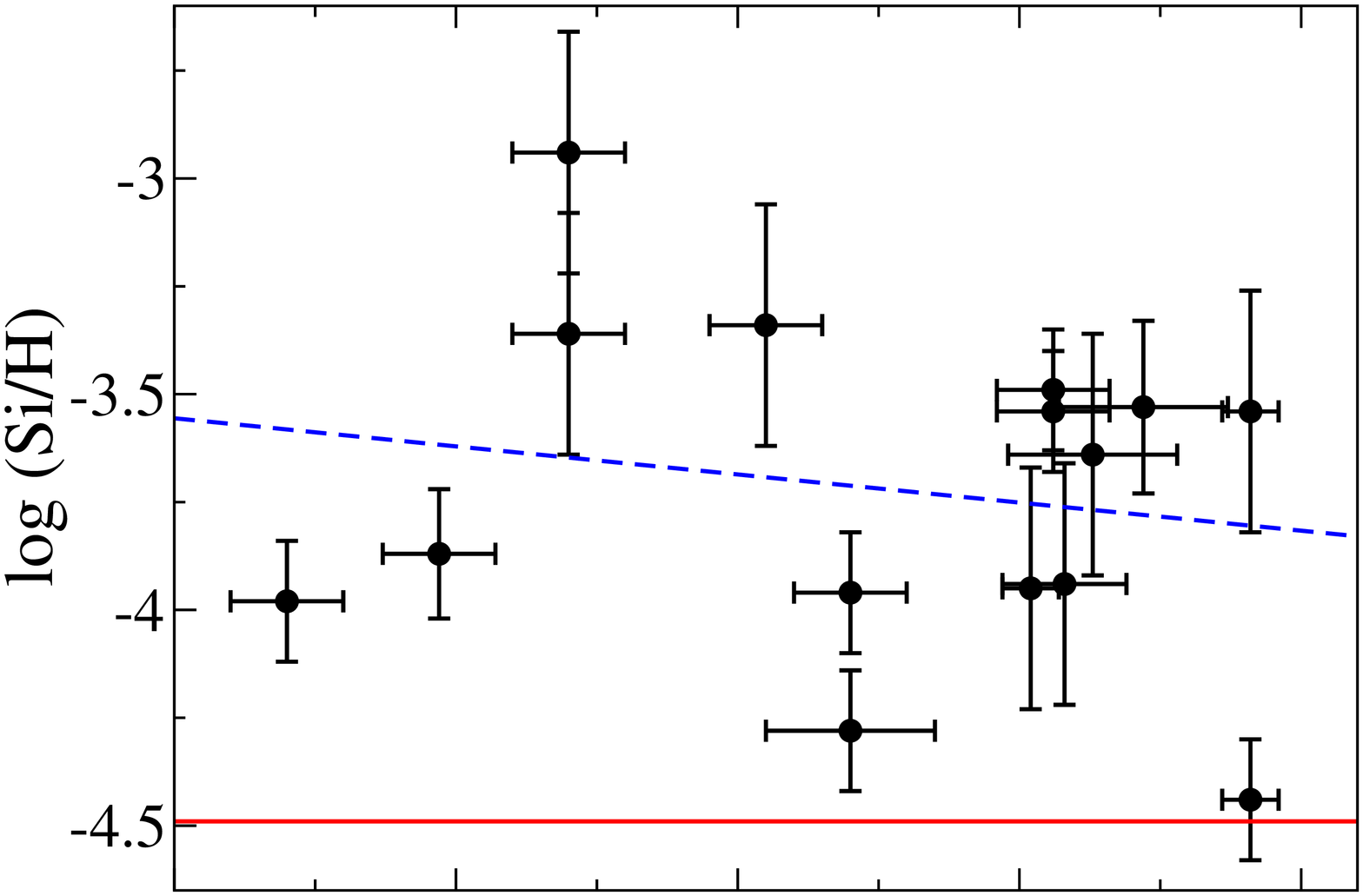}
\includegraphics[angle=-90,width=0.33\textwidth]{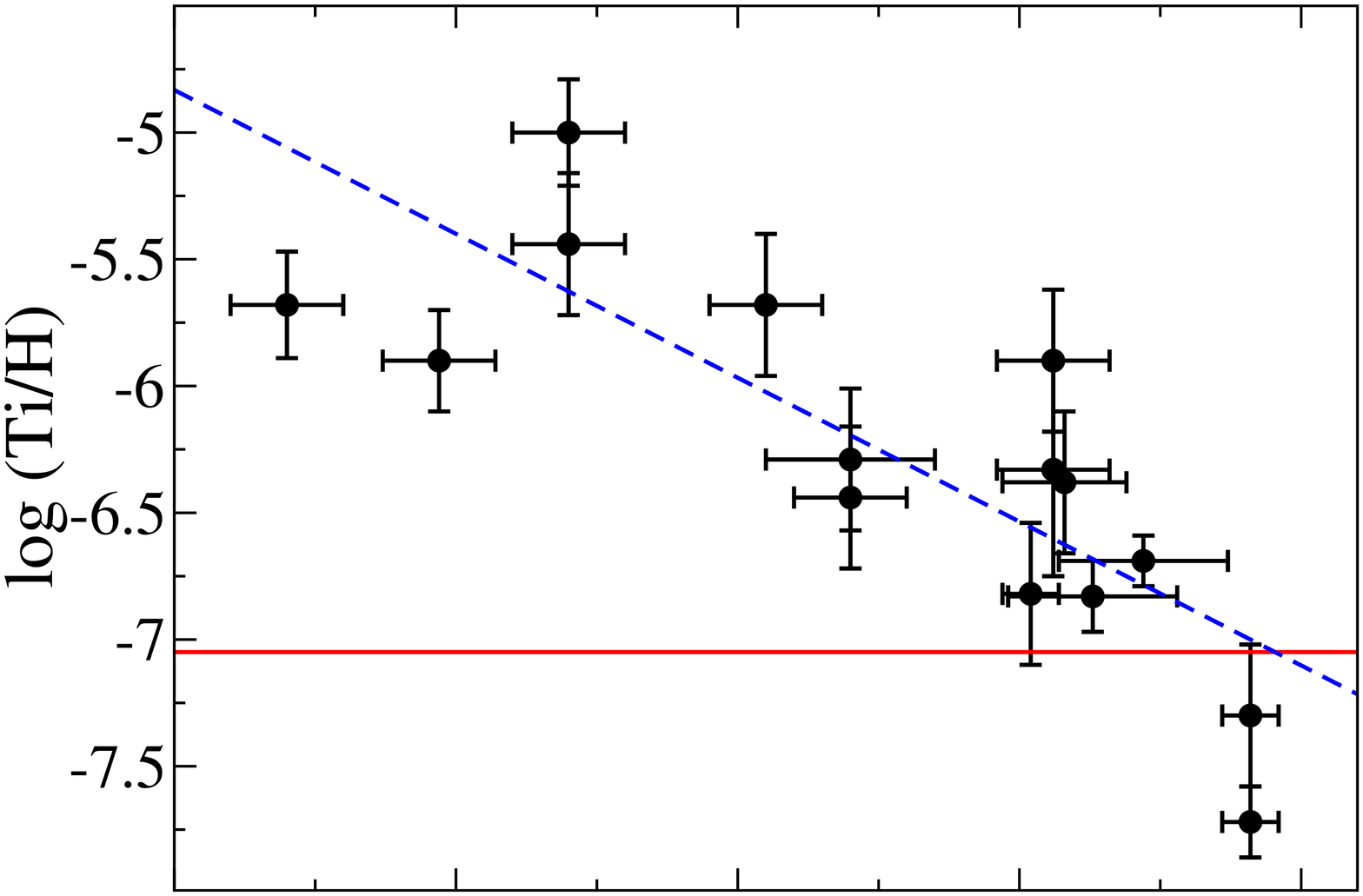}
\includegraphics[angle=-90,width=0.33\textwidth]{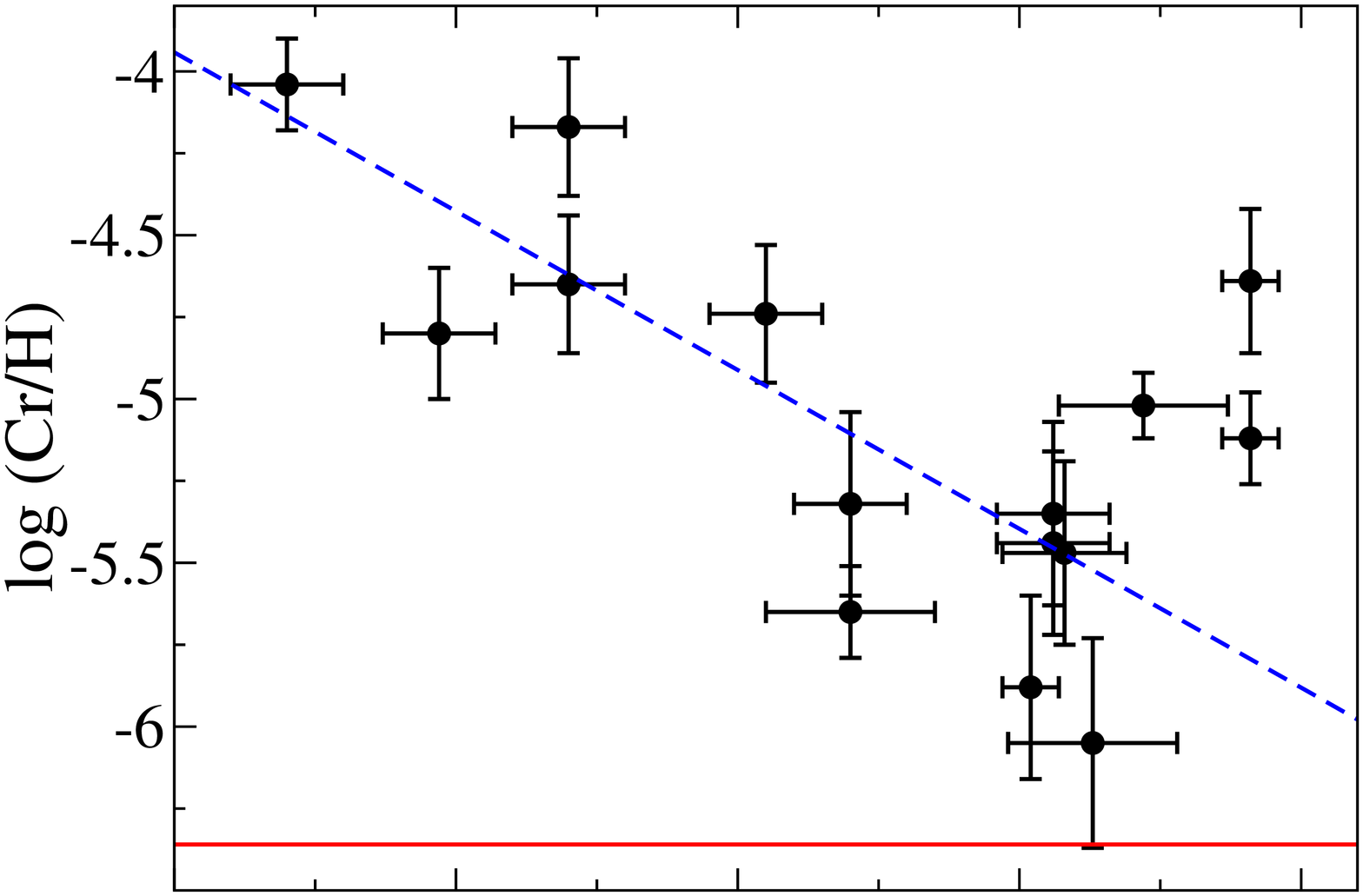}\\
\includegraphics[angle=-90,width=0.33\textwidth]{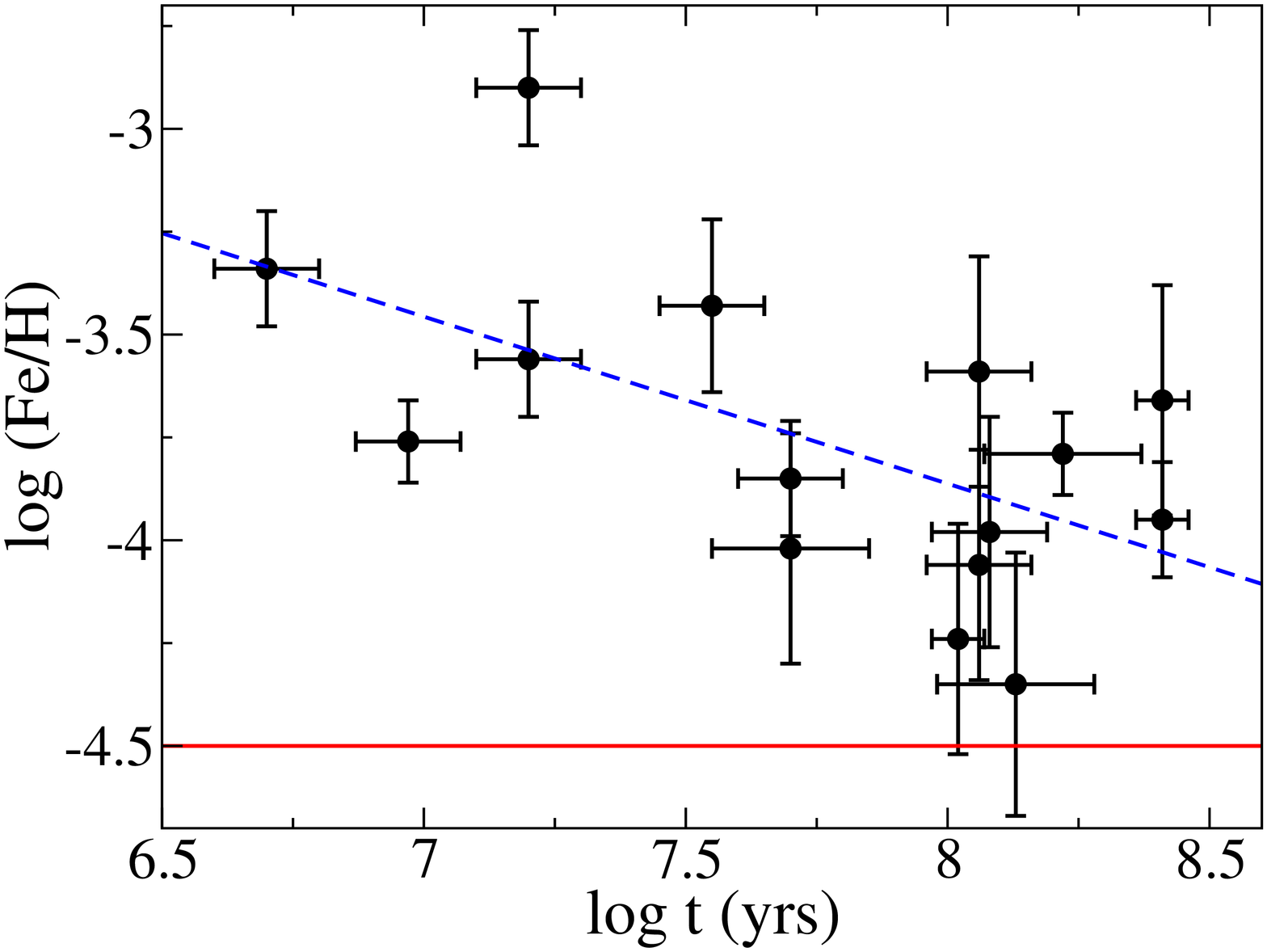}
\includegraphics[angle=-90,width=0.33\textwidth]{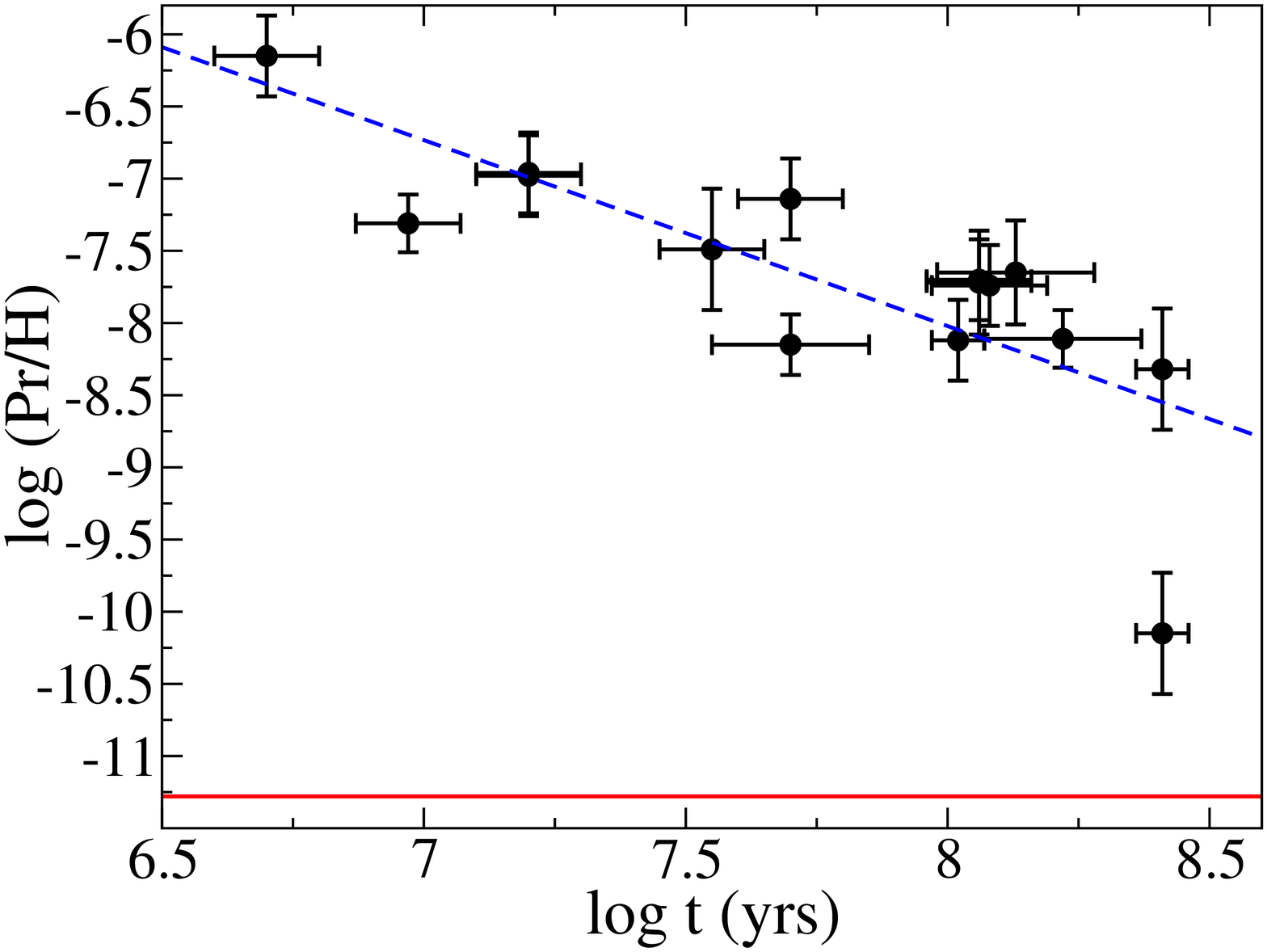}
\includegraphics[angle=-90,width=0.33\textwidth]{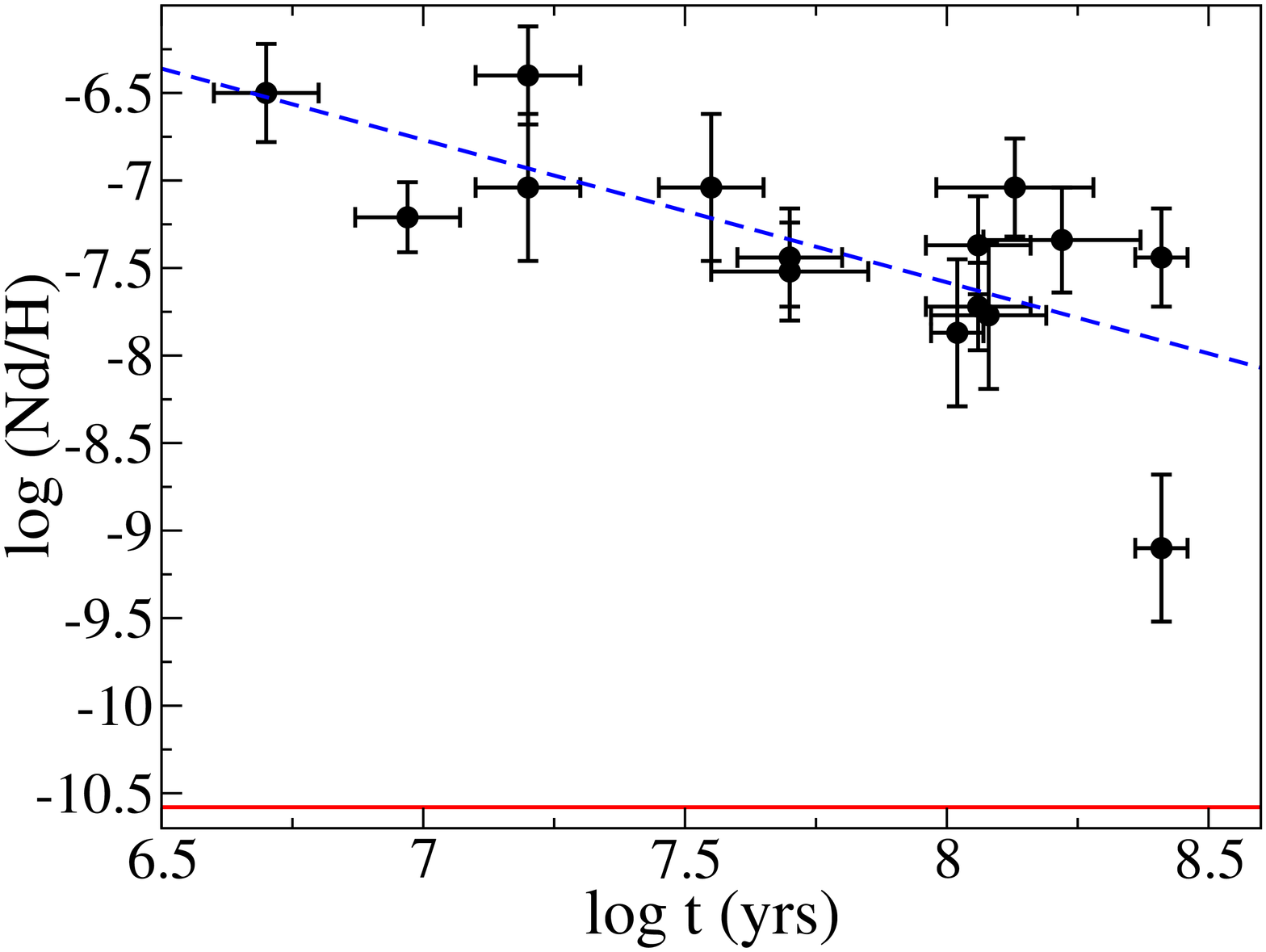}
\caption{Shown are the atmospheric abundances of He, O, Mg, Si, Ti, Cr, Fe, Pr, and Nd plotted against $\log t$. Each plot contains the solar abundance ratio for that element (solid red line) and the best-fit linear regression (dashed blue line).}
\label{abund-logt}
\end{figure*}

Figure~\ref{abund-logt} shows nine plots, one for each of the nine elements studied in this paper. In every panel we show the mean abundance found for each of the 15 stars of our sample as a function of the age of that star, $\log t$. Uncertainties in age are taken from Table~\ref{properties}, and in abundance from Table~\ref{abundances}. Although the error bars are large, in six of the nine panels a trend is clearly present. We can immediately see that on average several of the elements studied decrease or increase in abundance as our average $3.5 M_\odot$ star ages. It is essentially because we have good age resolution that we are able to detect these variations.

For each panel we show the best-fit linear regression to $\log N_{\rm X}/ N_{\rm H}$ versus $\log t$ (dashed blue line).  Table~\ref{lin-fits} tabulates the slope of the best-fit line with its uncertainty, as well as the significance of the slope, $\sigma_{D}$. This Table confirms the visual impression that the slopes are significantly (or nearly significantly) different from zero for He, the Fe-peak elements, and the rare earths, but not for the light elements: these elements evolve substantially in abundance through the main sequence life of magnetic Bp stars of about $3.5 M_\odot$.

We also show for comparison the solar abundance of each element studied (solid horizontal red line). It is clear that overabundance or underabundance relative to solar abundance is systematically present for all the elements studied. We discuss individual elements below.
\begin{figure*}
\centering
\includegraphics[angle=-90,width=0.33\textwidth]{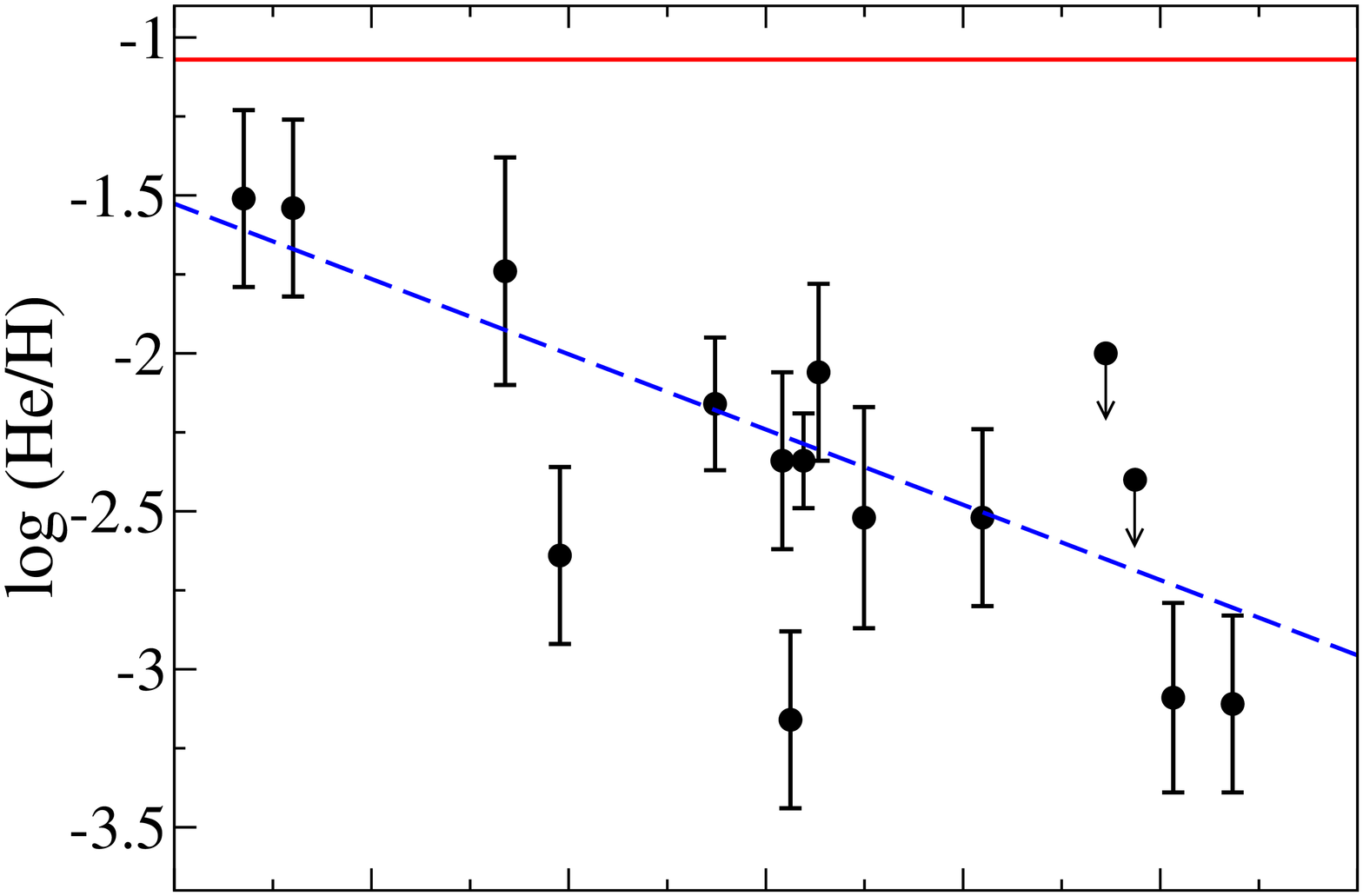}
\includegraphics[angle=-90,width=0.33\textwidth]{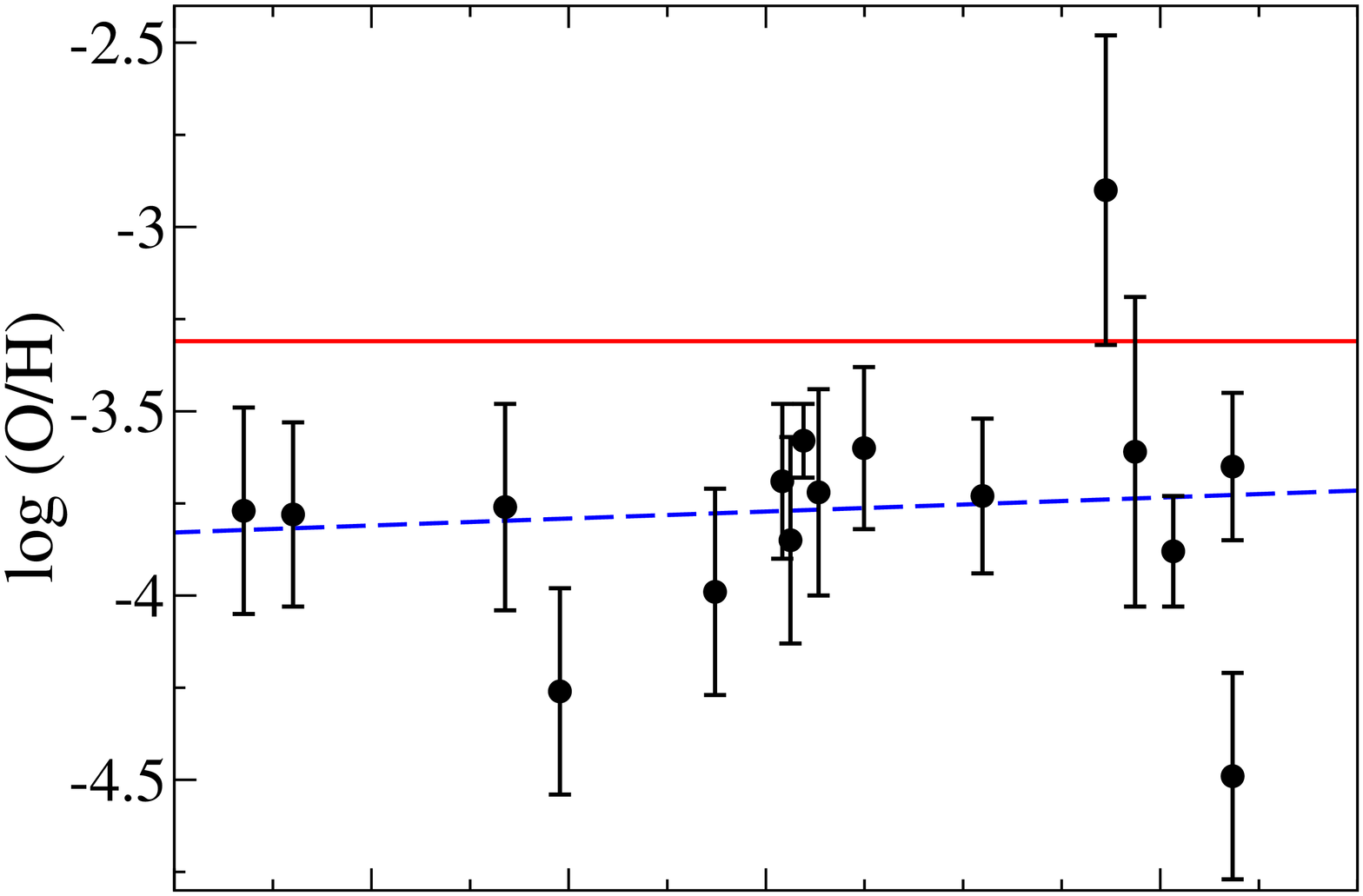}
\includegraphics[angle=-90,width=0.33\textwidth]{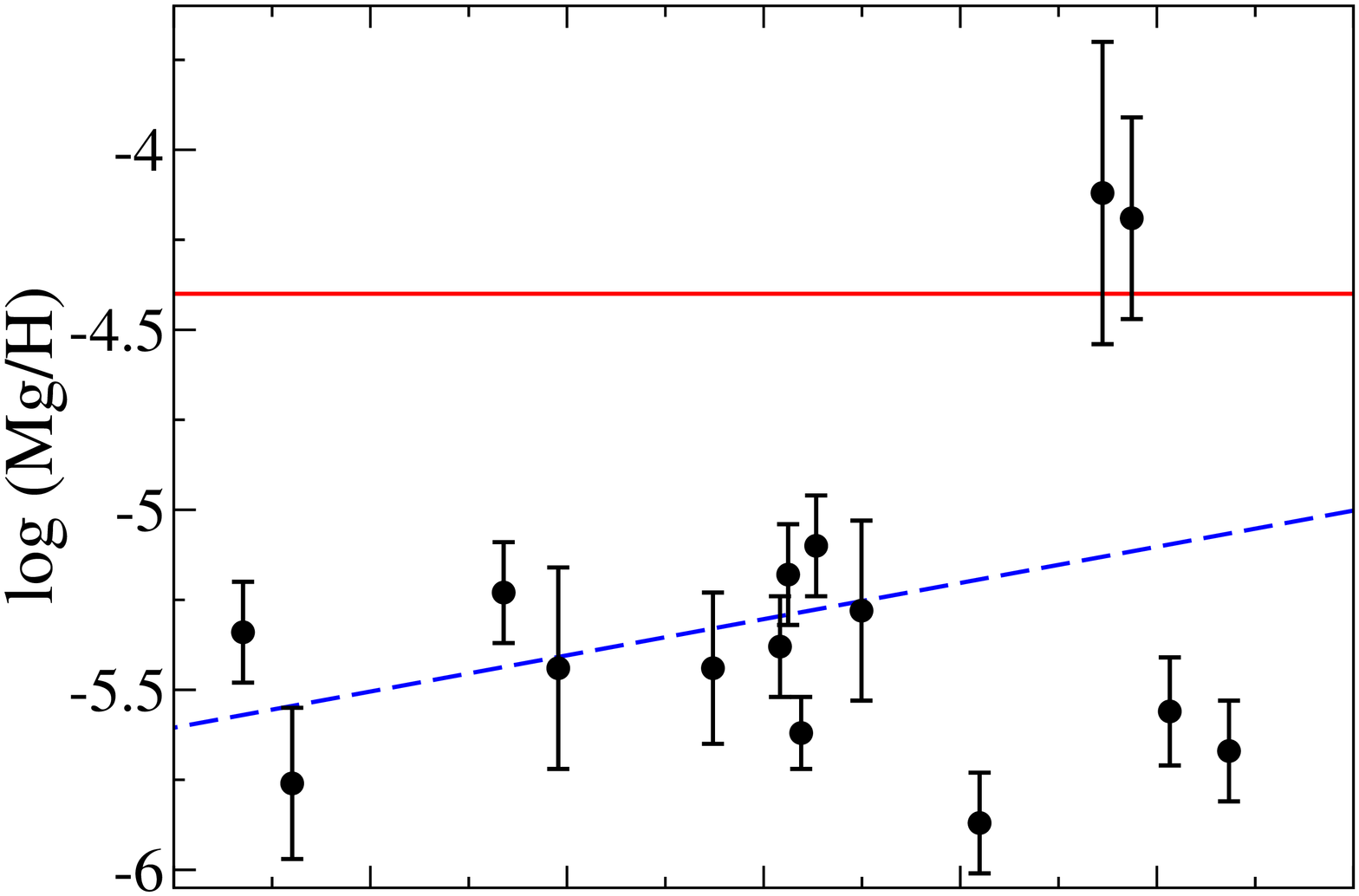}\\
\includegraphics[angle=-90,width=0.33\textwidth]{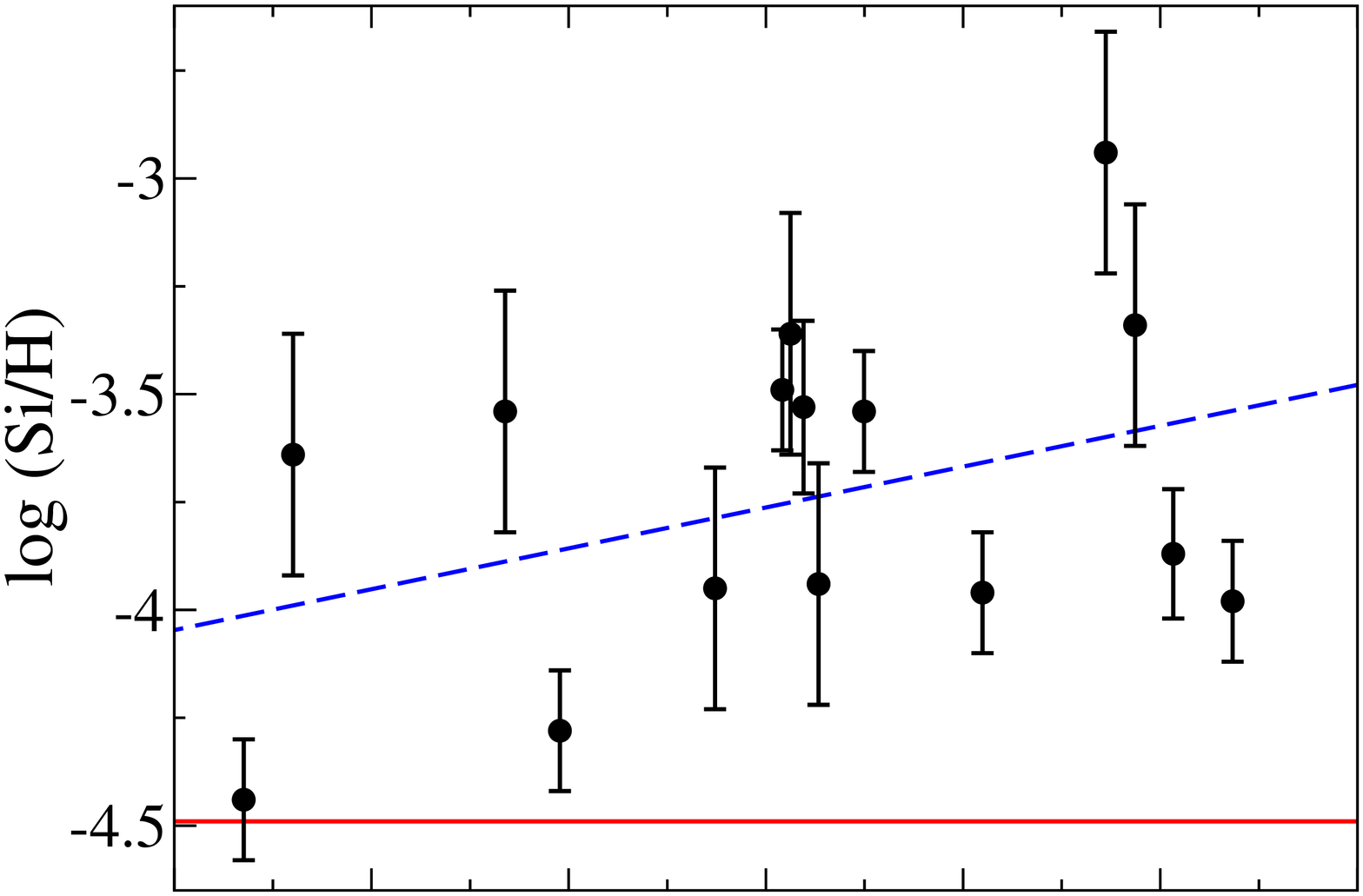}
\includegraphics[angle=-90,width=0.33\textwidth]{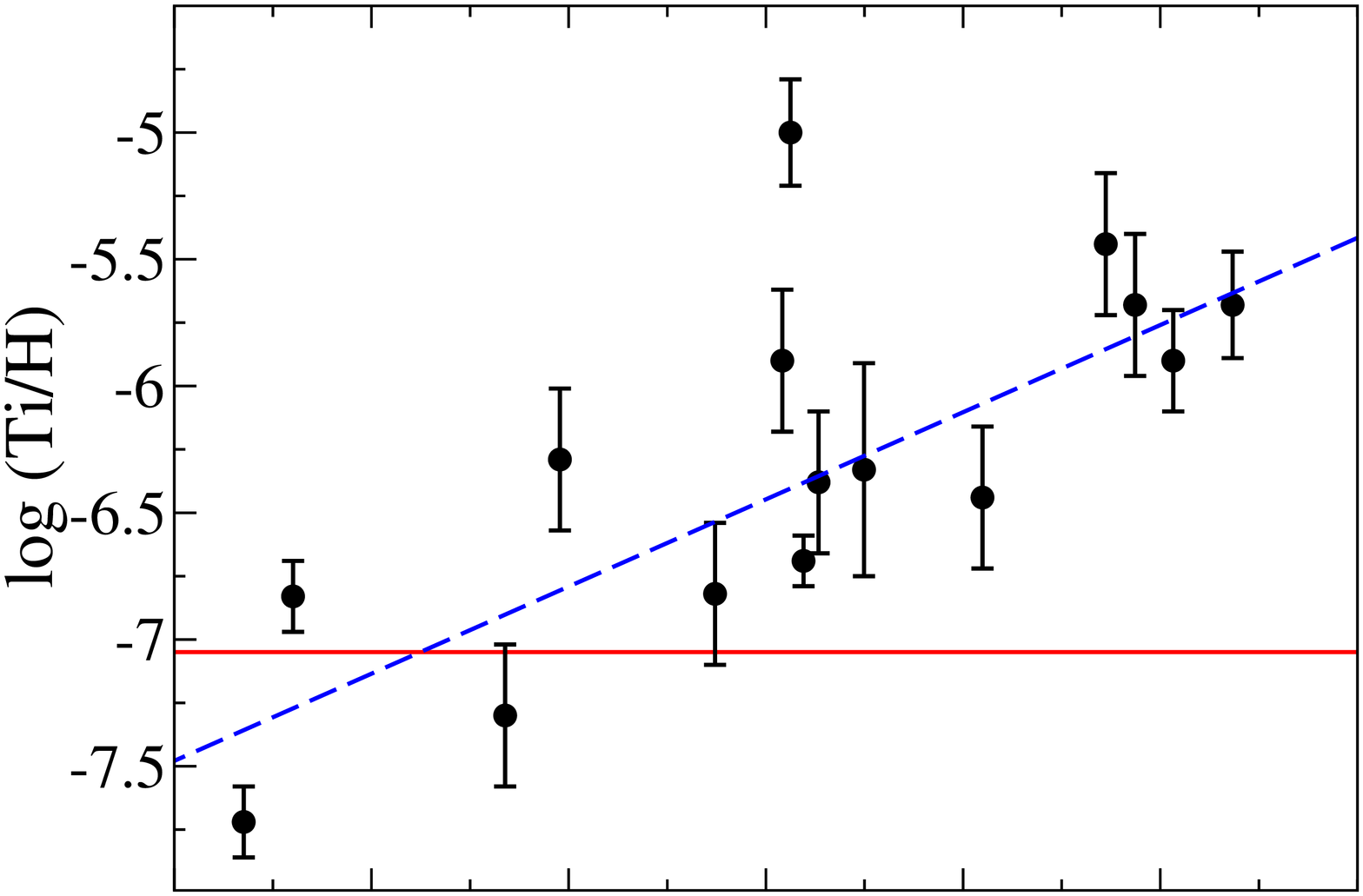}
\includegraphics[angle=-90,width=0.33\textwidth]{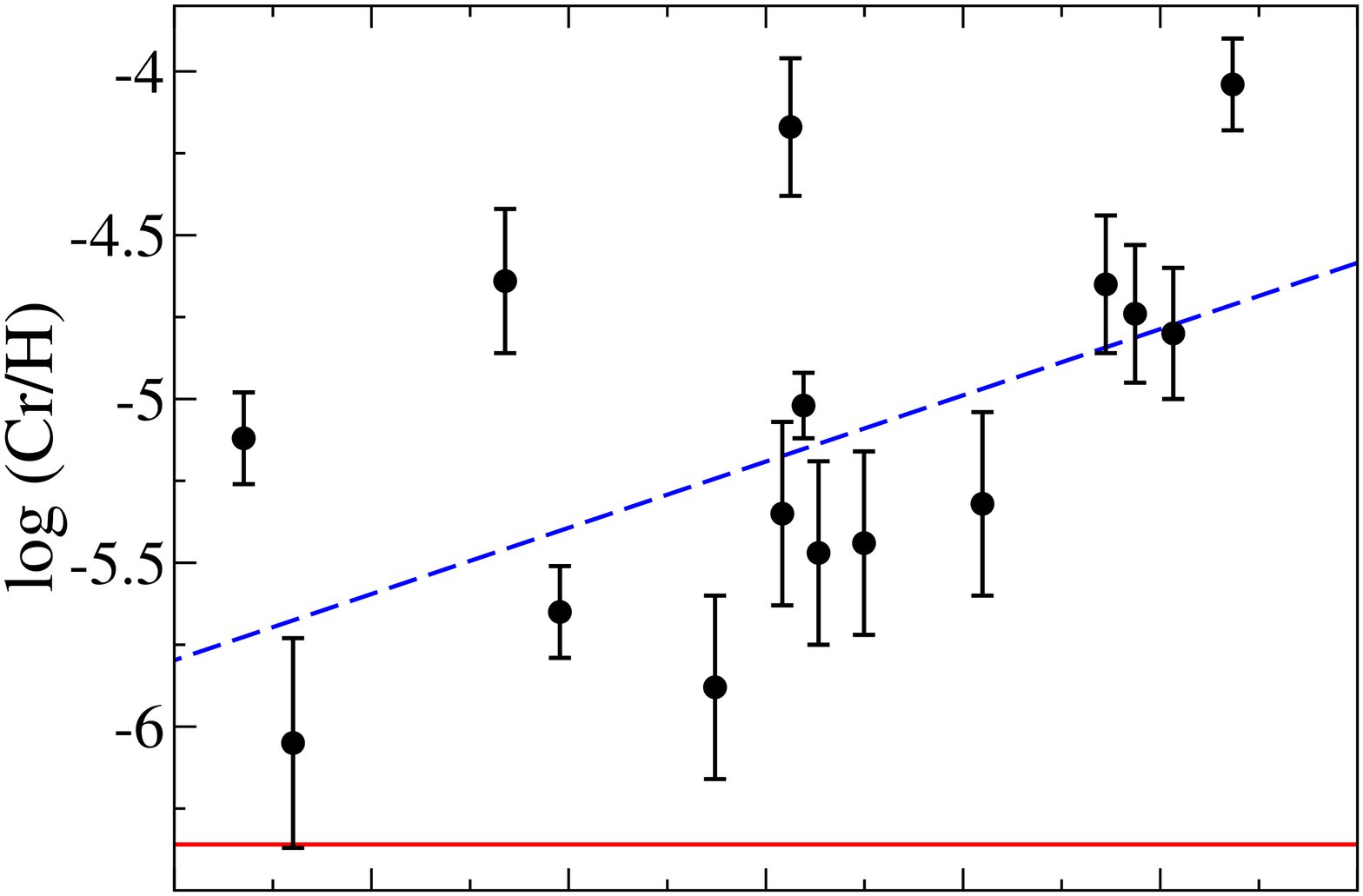}\\
\includegraphics[angle=-90,width=0.33\textwidth]{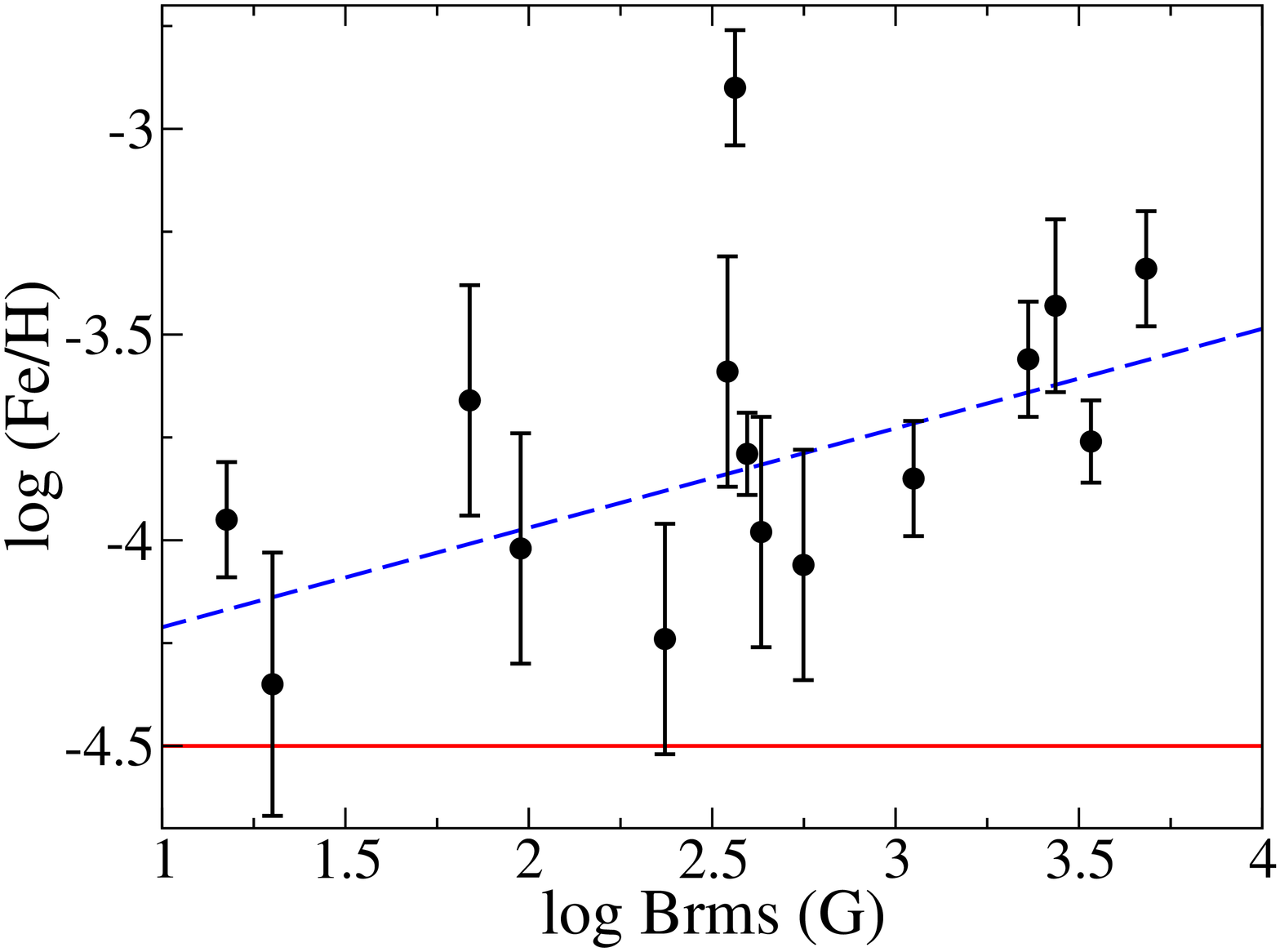}
\includegraphics[angle=-90,width=0.33\textwidth]{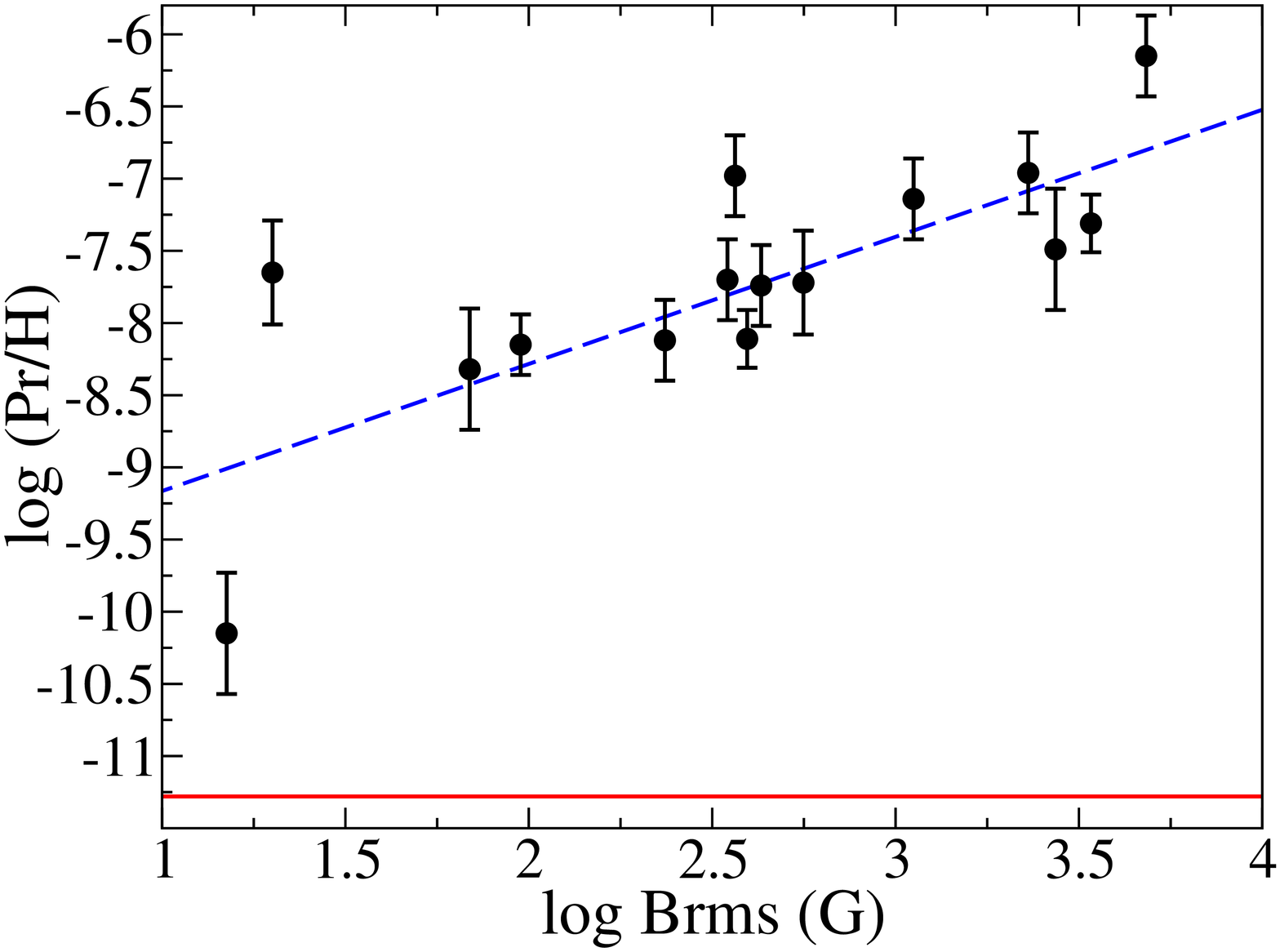}
\includegraphics[angle=-90,width=0.33\textwidth]{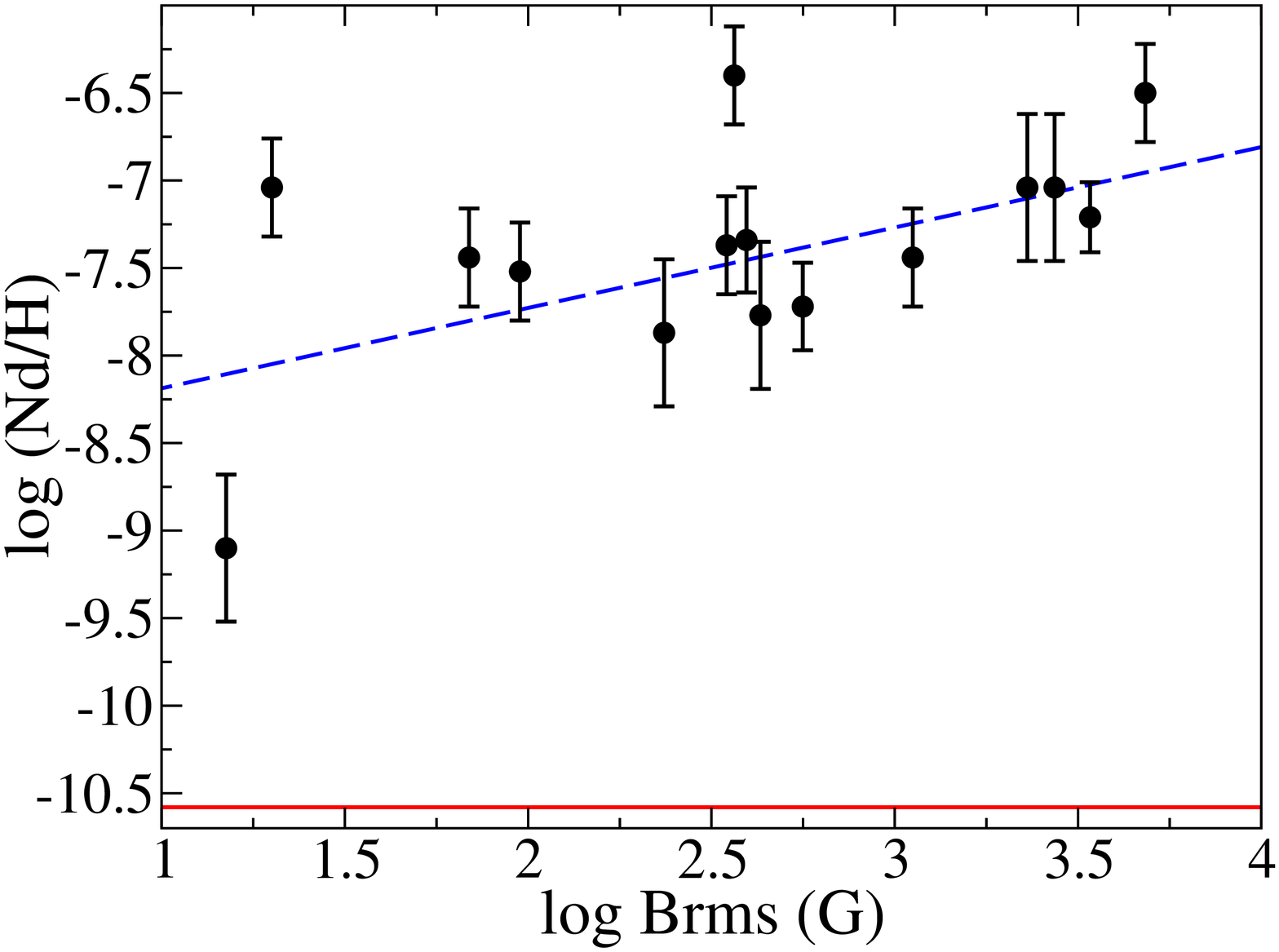}
\caption{Same as Fig.~\ref{abund-logt} but for $B_{\rm rms}$.}
\label{abund-brms}
\end{figure*}
\subsubsection{Helium}

One of the defining properties of magnetic Bp stars is the fact that they are visibly underabundant in He compared to normal stars of the same \te\ values.  This is true of all of the magnetic Bp stars in our sample. Unexpectedly, we observe a clear evolutionary {\em increase} in the abundance of He with stellar age. Substantial underabundances (of the order of 2~dex compared to the Sun) are found in stars near the ZAMS and those with ages up to about $\log t \sim 8$. Older stars in our sample (nearer the TAMS) show only modest underabundances of He, about 1 -- 0.5~dex smaller than the solar ratio.

\subsubsection{The light elements: oxygen, magnesium and silicon}

Oxygen, magnesium and silicon are systematically different from solar abundance ratios by values up to about 1~dex. No apparent trend in abundance  with age is seen in any of these three elements. Typically, both oxygen and magnesium are underabundant compared to the solar abundance ratios with  a few exceptions. For oxygen, only HD~133880 is overabundant compared to in the Sun; this anomalous behaviour is well-documented by \citet{Bailey2012}. HD~133880 and HD~45583 have nearly solar abundances of magnesium, nearly 0.5~dex larger than the other stars of the sample. It is not clear why these two (very similar) Bp stars depart so far from the general behaviour of other stars in our sample. In general, silicon appears to be overabundant compared to the solar abundance ratio throughout the main sequence evolution. Moderate Si overabundance is common for magnetic Bp stars between effective temperatures of about 10000 and 15000~K.

Although there are no clear trends in abundance for any of these three elements, all show a substantial star-to-star variations around the mean.

\subsubsection{The Fe-peak elements: Ti, Cr and Fe}

All three Fe-peak elements studied here, Ti, Cr and Fe, are quite overabundant compared to their solar ratios early in the main sequence phase. However, the level of excess abundance is larger for the lower abundance elements Ti and Cr than for Fe. In fact, Cr reaches almost the same level of total abundance as Fe close to the ZAMS. For these elements, definite decreases in atmospheric abundance are observed with time and by the end of the main sequence all three elements are not much more abundant that the solar values. 

\subsubsection{The rare earth elements: Pr and Nd}

Close to the ZAMS, both these elements appear very overabundant (more than 4~dex) relative to the Sun. There is a clear decrease in the abundances of each element with age, although even at the TAMS both elements are still typically 2--3~dex overabundant. The apparently smaller scatter around the regression lines compared to other elements is produced mainly by the very large range seen in the y-axis of these plots.

The single outlier with low abundance near the TAMS is HD~162576. This star also has the smallest Si abundances.
\begin{figure*}
\centering
\includegraphics[angle=-90,width=0.85\textwidth]{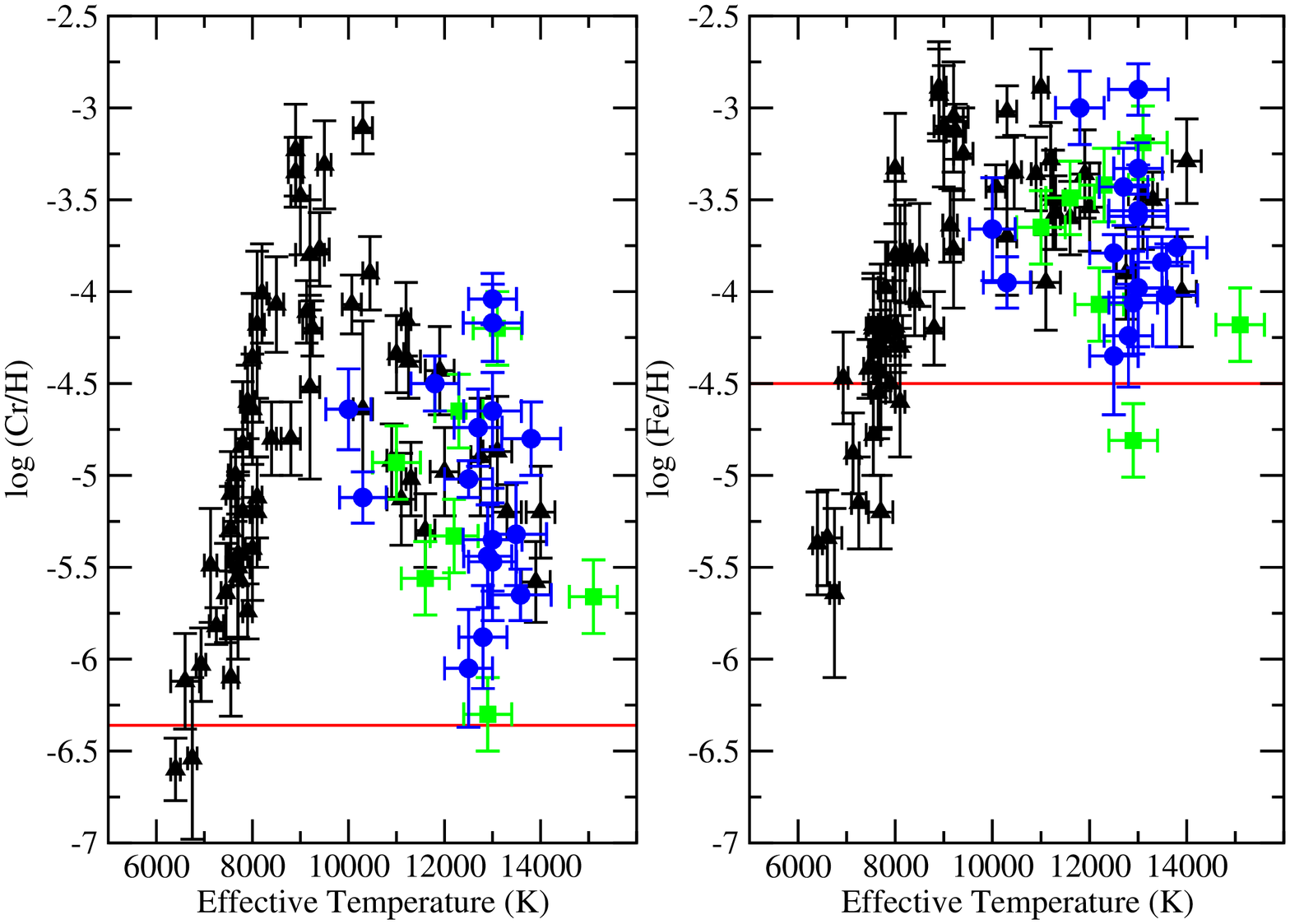}
\caption{Shown are Cr (left) and Fe (right) abundances verses \te. The black triangles are data taken from \citet{Ryabchikova2004} and \citet{Ryab2005}, the green squares are from \citet{BL2013} and the blue circles are data from this paper.}
\label{cr-fe-teff}
\end{figure*}

\subsection{Abundance variations with magnetic field strength}
\citet{paper2,paper3} discovered that magnetic field strength decreases during the main sequence lifetime of magnetic Bp stars. Since we see trends in abundance with time for He, Ti, Cr, Fe, Pr and Nd, we expect that there will be correlations between abundances and stellar magnetic field strength. Fig.~\ref{abund-brms} plots abundances versus $ \log B_{\rm rms}$ in the same manner as for $\log t$. In all the elements for which a trend was observed versus time (He, Ti, Cr, Fe, Pr and Nd) a correlation is also found for abundance versus $\log B_{\rm rms}$, but in the opposite sense: helium abundance decreases with increasing magnetic field strength and the Fe-peak and rare-earth elements increase in abundance with increasing magnetic field strength. It is clear that the correlations seen in Figure~\ref{abund-brms} are those expected from the known decline of \bz\ with age together with the newly discovered variations of abundance with age. However, the observed correlations may also contain some information about how the time evolution of atmospheric abundance is modified by magnetic fields of various strengths. This may become clear with more detailed modelling. 

We note that the results shown in Figure~\ref{abund-brms} seem to be the explanation of the correlation between the Geneva photometric measurements of the 5200~\AA\ depression (using the Z index) and field strength, as discussed by \citet{Cramer1980}. This effect has also been exploited recently by the Special Astrophysical Observatory (SAO) group \citep[e.g.][]{Kudry2006,Kudry2008,Kudry2011} to search for particularly large magnetic fields in Ap/Bp stars.

\subsection{Cr and Fe abundances versus effective temperature}

Our results can be compared to previous studies of magnetic Bp stars. \citet{Ryabchikova2004} presented a comprehensive comparison of Cr and Fe abundances in roAp (rapidly oscillating Ap) stars versus effective temperature. However, their study was mostly restricted to stars at or below around 10000~K. \citet{Ryab2005} added stars between about 10000 and 15000~K and our current study, as well as a previous study by \citet{Bailey2013}, adds 23 stars in that same temperature range. The results are shown in Fig.~\ref{cr-fe-teff}. Our observations confirm the tendency, already visible in the data of Ryabchikova, for the star with \te\ above 8000~K to exhibit substantially larger abundance dispersion than is shown by cooler stars. Our new results suggest that this extra scatter might be due to mixing stars of quite different ages at a given value of \te, although we do not yet have enough abundance measurements of magnetic stars with both known mass and known age to confirm or reject this idea. It appears that for Cr the largest observed abundances peak at around 10500~K and decline with increasing effective temperature. A similar trend is seen with Fe, but to a lesser extent. There is only a broad flat peak in the abundance of Fe with \te. These results support the conclusions of \citet{Ryabchikova2004} that the maximum abundance for Cr and Fe both approach the same value of between $-3$ and $-4$~dex.

\section{Discussion}
\begin{figure*}
\centering 
\includegraphics[angle=-90,width=0.9\textwidth]{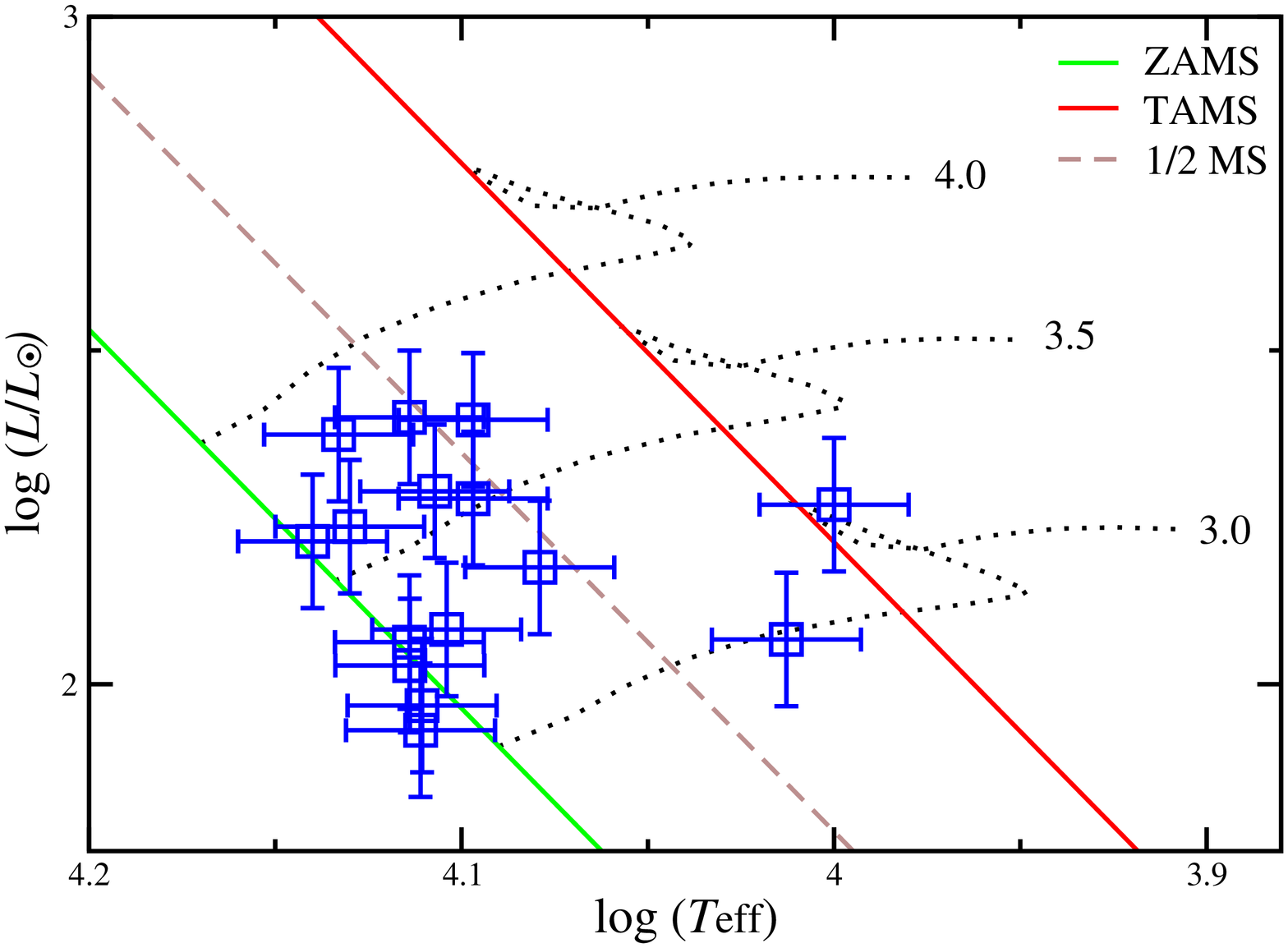}
\caption{The position of the stars of this study in the
  Hertzsprung-Russell diagram, compared to the main sequence evolution
  tracks for stars with 3.0, 3.5 and 4.0M~$_{\odot}$. Also shown are
  the ZAMS, the TAMS, and the points where half the main sequence lifetime is
  completed.}
\label{HRdiag}
\end{figure*}
It is well known that during its evolution on the main sequence, a star undergoes an extensive process of climate change. The atmospheric effective temperature decreases by about 30\,\%, the radius increases by a factor of order three, and the gravity decreases by a factor of ten. In effect, by studying the changes in average atmospheric chemical composition of stars in our sample, with its limited mass range, as a function of cluster age, we are observing the time evolution of the atmospheric chemistry of a $3.5 M_\odot$ star through its main sequence life. Figure~\ref{HRdiag} shows a Hertzsprung-Russell (HR) diagram for the stars in this study, compared to theoretical stellar evolutionary models \citep{Giretal00}.

What we have observed, for the first time, is a clear evidence of the secular evolution of the atmospheric chemistry of the magnetic peculiar B-type (Bp) stars during the main sequence phase. That is, we have observed how the consequences of diffusion processes are modified by time and by stellar climate change. Figure~\ref{abund-logt} shows that the abundances of He, Ti, Cr, Fe, Pr, and Nd change monotonically with time. In contrast, the light metals O, Mg and Si also have peculiar abundances, but show no significant trends with age.

It is generally believed that the peculiar chemical abundances found in the atmospheres of magnetic Ap and Bp stars are the result of microscopic diffusion, in competition with other processes such as turbulent diffusion, convection, meridional circulation, well-mixed mass loss, accretion, and the effects of a magnetic field. 

The basic ideas of how diffusion can lead to anomalous atmospheric abundances have been discussed for some years \citep{Michaud1976,Vauclair1983}. Essentially, in a quiescent stellar plasma, the gravitational field tends to cause trace atoms to diffuse downward relative to the dominant hydrogen gas, but the outward flow of radiation from centre to surface is
responsible for a force which tends to levitate some trace atoms up through the atmosphere, if they absorb at many wavelengths and are not too abundant. In slowly rotating middle main sequence stars, which have relatively weak mixing near the surface, this phenomenon can lead to anomalous atmospheric chemical abundance.

Detailed and quantitative modelling of diffusion is actually quite complicated. The average upward acceleration per ion of a particular species depends on the intensity of the radiation (which increases with effective temperature \te), on the specific atomic transitions that the ion can undergo (especially those arising from low-lying energy levels), and on the number density of the trace ion. In general, as the fractional number density of the trace ion increases, the acceleration per ion decreases. For many trace ions of low abundance (having, say, number density less than about $10^{-6}$ that of H) in stars of \te\ above about 7000~K, the upward acceleration $g_{\rm R}$ due to the radiation field is larger than the local acceleration of gravity $g$, and the ion diffuses upwards rather than downward. In slowly rotating middle main sequence stars, which have relatively weak mixing near the surface, this phenomenon can lead to anomalous chemical abundances in the stellar atmospheres.

The actual vertical variation of radiative acceleration for various ions in specific stellar models has been calculated by a number of groups. The calculations fall into two general types. Because the acceleration per ion decreases as the fractional abundance of the ion increases (due to saturation of the spectral lines that the ion absorbs), an ion which is levitated at low abundance can become locally abundant enough that $g_{\rm R} = g$, a situation in which the tendency of that trace ion to diffuse upwards or downwards vanishes. It is possible to determine the fractional abundance for a specific atom as a function of position in a region (e.g. with height in the stellar atmosphere) for which this condition is satisfied everywhere in the region, and diffusion of this atom ceases. The run of abundance through the region satisfying this condition is known as the ``equilibrium abundance distribution''. Some recent computations of equilibrium abundance distributions in the atmospheres of the late B stars are re reported by \citet{LeBlanc2003}, \citet{LeBlanc2009} and \citet{Stift2012}.

Recent work in this field has focussed on re-computing the structure of atmospheres in which equilibrium abundance distribution computations suggest the presence of strong vertical variations in abundance, in order to have the atmospheric structure be consistent with the abundance stratification \citep{LeBlanc2003,LeBlanc2004,Shulyak2009,LeBlanc2009,Stift2012}. It is found that when large variations in abundances of abundant elements with altitude are present, the atmospheric structure is substantially perturbed.

However, an equilibrium abundance distribution may not be achievable in a region if not enough atoms of the species is available from below. In this second case, atoms may diffuse slowly up into the region from below and at the same time be removed from the top. This situation could settle into a stationary state in which the flux of atoms through the region is constant with height, and the abundance distribution is unchanging but the actual abundance is smaller than the value leading to equilibrium. In principle, if a large enough volume of the star is considered, the evolution of abundance of an atom with radius, including through the atmosphere, could be computed as a function of time. Because the time-scale for evolution varies strongly with density, this is a stiff problem. Solutions on the scale of a stellar envelope, to try to explain the chemical abundances in the atmospheres of metallic-line (Am) stars, including either a (mixed) stellar wind or deep, but weak, turbulent mixing have been reported by \citet{MRV2011} and \citet[][earlier work on abundance evolution is cited there]{Vick2010}. 

The time variation with height of abundance of a very unabundant element diffusing up through a stellar atmosphere has recently been studied by \citet{Alecian2011}, who find that a stationary state of constant particle flux with height is generally achieved after a few hundred years, provided that the atoms continue to diffuse out of the top of the atmosphere. The actual value of the flux, and the run of abundance with height, is set by the abundance of the atom supplied at the bottom of the atmosphere from the reservoir below. We will refer to this situation as ``stationary flow-through''. 

We now apply these general ideas to the elements studied here, in stellar atmospheres of $\log g \sim 4$ and $\te \sim 10-13\,000$~K. In the panels of Figure~\ref{abund-logt}, it appears that we can identify several different cases. 

\subsection{Abundant light elements: He, O}

The general behaviour of these elements was already predicted by \citet{Michaud1976}. For the very abundant light elements He and O, the radiative acceleration for $\te \sim 12000$~K is much too weak to support a solar abundance of these elements (cf. $g_{\rm rad}$ calculations for $\te = 12000$~K by \citealt{Hui2002}). These two elements are expected to diffuse downward into the stellar envelope below the atmosphere until the relative abundance has fallen enough that radiative acceleration can support them in the atmosphere. Consequently, He and O are expected to have abundances well below the solar values, and this is observed.

The diffusion of He in stellar atmospheres with well-mixed winds (including H) has been studied in more detail by \citet{Vauclair1991}, \citet{Krticka2004} and \citet{Theado2005}; all groups confirm that without a mixed wind from the star, that He should diffuse downward until quite low relative abundance is reached. However, we know of no published predictions of how low the He abundance should drop before an equilibrium distribution is achieved. It may well be that some degree of turbulent mixing with sub-atmospheric layers is required to keep the abundance of He as large as is observed (Figure~\ref{abund-logt}).

The {\em increase} in He abundance with stellar age that we observe is particularly unexpected. Radiative levitation is so weak (due to the shadowing of resonance lines of He~{\sc i} by the very strong H continuum bound-free absorption below the Lyman limit) that it is not expected to play a significant role in the observed He abundance, and especially not in its increase with decreasing \te. The observed increase of He abundance towards the solar mixing ratio suggests either that there is significant and increasing mixing upward of envelope He by some process unrelated to diffusion, or that there is accretion of (He-rich) interstellar gas.

\citet{Landstreet1998} have studied the diffusion of O in stars in our mass range, using a simple approximation to estimate the radiative acceleration. Extrapolating their results, it appears that radiation may be able to support O at an abundance one to two dex below the solar abundance. This conclusion is supported by the very small $g_{\rm rad}$ value for O at $12000$~K found by \citet{Hui2002}. The rather mild underabundance of O that we observe (about 0.5~dex) is probably higher than the level that would be found by an equilibrium abundance calculation. Furthermore, as the stars in our sample age, \te\ and $\log g$ both decrease. The change in \te\ typically reduces $g_{\rm R}$. However, the decrease in $\log g$ means that a smaller value of $g_{\rm R}$ is required to support ions of O, which in turn means that a larger abundance can be supported. It is not clear which of these two effects dominates, and in fact the observed lack of significant variation in the O abundance with age suggests that if the observed abundance is the result of radiative levitation, the two effects roughly cancel. We do not have any explanation at present for the slight overabundance of O observed for a single star, HD~133880.

It would clearly be of interest to have available some published results of computed equilibrium abundances of He and O in our temperature range, and especially to have such calculations follow the evolution of a star of $3.5 M_\odot$ from $\te = 13000$~K and $\log g = 4.5$ to $\te = 10000$~K and $\log g = 3.5$, in order to determine the importance of radiative levitation of He and O in the observed abundance evolution of the stars of our study.

\subsection{Light metals: Mg and Si}

The Mg abundance, based generally only on the 4481~\AA\ line of Mg~{\sc ii}, appears to be about 1~dex below the solar abundance, or $\log(n_{\rm Mg}/n_{\rm H}) \approx -5.2$. This may be compared with the predicted Mg abundance profile for a star of $\te = 12000$~K of \citet{LeBlanc2009} and of \citet{Alecian2010}. The abundance predicted by the equilibrium calculation ($g_{\rm R} = g$ throughout the atmosphere) is mildly non-uniform through the line-forming region, but is of about this same values. Thus it appears that Mg may be an element in which the supply of atoms from below has been large enough to allow the development of an equilibrium stratification.

No strong trend of abundance with age is observed. We have no explanation at present for the two stars (HD~133880 and HD~45583) that deviate strongly from the mean behaviour, with Mg abundances about 1~dex larger than other, younger and older, stars. 

Since Mg may well be described in the stars of our sample by the equilibrium abundance distribution, it would be of great interest to have calculations of the equilibrium atmospheric abundance of Mg following the evolution of \te\ and $\log g$ for our $3.5 M_\odot$ stars.
 
Silicon equilibrium atmospheric abundance has been studied by a number of authors, including \citet{Alecian1981}, \citet{LeBlanc2009}, and \citet{Alecian2010}. While detailed results differ, the overall conclusion is that Si is expected to be of order 1~dex underabundant in the atmosphere at $\te \sim 12000$~K. Instead, as has been found in the past, we observe an overabundance of about 1~dex at all ages. This situation has been a long-standing puzzle. Because it is not clear how to obtain an atmospheric abundance that is nearly 2~dex larger that the maximum value that can be supported by radiation pressure, various explanations have been probed, such as support by a horizontal magnetic field, non-LTE effects, etc., but none have been found to offer convincing explanations of the observed overabundance at all ages. 

\subsection{Iron peak elements: Ti, Cr and Fe}

Expected abundances for these elements based on equilibrium have recently been computed for stars in our mass range by \citet{LeBlanc2009}, \citet{Alecian2010}, and (for Fe) by \citet{Stift2012}. The results are most extensive for iron. The computed equilibrium abundances are only qualitatively in agreement with one another, and depend on still uncertain physics, particularly on how to treat redistribution of momentum between ionisation stages, but also on the effects of the magnetic field, and the chemical composition  assumed for the computation. However, all the computations of equilibrium abundances agree that Fe in the line forming region of stars in our mass range should reach equilibrium at an abundance of around $-3$ to $-3.5$~dex relative to H. This is reasonably consistent with the values that we observe (Figure 2), so this may well be an element for which the equilibrium assumption of zero diffusion is approximately correct. This implies that an adequate supply of Fe ions is available from below the atmosphere to replenish atmospheric atoms lost to space during the initial period of equilibration, and to keep the abundance high enough to satisfy the equilibrium condition.

It is clear from the few values of \te\ for which equilibrium Fe abundances have been computed that the radiative acceleration and the equilibrium abundance decrease with decreasing effective temperature. However, there is no computational information on how the equilibrium abundance varies as $\log g$ decreases from 4.5 to 3.5, except for the qualitative result that the effect of the decrease in $\log g$ should be to make it possible for a given radiative force to support more atoms. Since the observed evolution of Fe abundance is that the abundance decreases with time, it appears that the decrease in radiative acceleration with decreasing \te\ may dominate. This is certainly a question that could be studied by computing an appropriate series of equilibrium abundance models following the evolution of a $3.5 M_\odot$ star

The available equilibrium calculations for Ti and Cr in the mass range of our observations by \citet{LeBlanc2009} have been cut off in the line-forming region because of artificial limits of 1000 times the solar abundances imposed on the calculations. Thus we may suppose that without these limits, the equilibrium calculations would imply equilibrium abundances of Ti and Cr of $\log(n_{\rm el}/n_{\rm H})$ similar to those of Fe, in the range of $-3$ to $-4$. This is confirmed by the calculations of \citet{Alecian2010}. Although we observe overabundances of these two elements of up to 2~dex when the stars are young, the Cr abundance seems to be generally less than the equilibrium value, and the Ti abundances almost certainly are below equilibrium even at young ages. These may well be elements for which the abundance would be closer to that predicted by the assumption of a
stationary flow-through state, with atoms fed into the atmosphere from below and lost from the top \citep[cf][]{Alecian2011}. In this case the limiting factor determining the atmospheric abundances is the available number density of Cr or Ti brought up from the envelope by diffusion.

As for iron, it is not known whether the probable decrease in radiative acceleration with declining \te, or the increased abundance that can be supported as $\log g$ decreases, is the dominant effect during main sequence evolution. It appears that the effect of decreasing radiative acceleration, which may well reduce the supply of Cr and Ti at the bottom of the atmosphere and thus may reduce the supply if these elements are in a stationary flow-through state but not in equilibrium, may dominate, as the abundances of both these elements are observed to decrease strongly with increasing stellar age \citep[cf][]{Alecian2010}. It would be of great interest to have computations of equilibrium abundances of Ti and Cr following the evolution of a $3.5 M_\odot$ star for more direct comparison with our results.

\subsection{Trace heavy elements: Nd and Pr}

There are no calculations of the equilibrium abundances of heavy elements such as Nd and Pr near the \te\ temperature range of interest to us here, except for the equilibrium
stratification calculated for a simplified artificial low-abundance element with $\te = 12000$~K by \citet{Alecian2011}. This artificial element is expected to behave somewhat like Hg, and the equilibrium abundance computed for it is about $-8$~dex without a magnetic field, and possibly even smaller high in the atmosphere in the presence of a strong magnetic field. It is not clear to what extent this result is applicable to Nd or Pr. More realistic computations would clearly be of great interest. 

In considerably cooler magnetic Ap stars, it is known from modelling of observed spectra that these elements are strongly stratified, with the abundance of Nd as much as 4~dex more abundant high in the atmosphere than near optical depth unity \citep{Mashonkina2004,Nesvacil2013}. However, we know of no similar stratification models of rare earths in Ap stars in our temperature range. The stars in our sample are hot enough that we have not been able to identify the Nd~{\sc ii} lines that would make stratification analysis possible. 

We thus have no real evidence as to whether Nd and Pr at the temperatures of our stellar sample have approximately equilibrium abundance distributions, or are in a stationary flow-through state in which these elements are entering the bottom of the atmosphere from below and are being lost to space from the top.

Like the Fe-peak elements, these rare earths are observed to have mean abundances that clearly decrease with stellar age. This could be because the radiative acceleration upward is {\bf probably} declining with decreasing \te, and thus either decreasing the equilibrium abundance in the atmosphere, or decreasing the supply at the bottom of the atmosphere if the flow-through case applies.

\section{Summary and conclusions}
In this paper, we report the discovery of the time variations of atmospheric abundances during the main sequence lifetime of a magnetic Bp star. Large overabundances are observed for Ti, Cr,  and Fe near the ZAMS. As the star evolves on the main sequence, the abundances of these Fe-peak elements clearly decrease, approaching nearly solar values closer to the TAMS. The rare-earth elements Pr and Nd show drastic overabundances in young Bp stars, with values nearly 10$^{4}$ times greater than in the Sun. Near the TAMS, magnetic Bp stars remain overabundant in rare-earth elements, still exhibiting abundances of Pr and Nd that are at least a factor of 100 larger than the solar ratios. The light elements O, Mg and Si show no evidence of time variations. However, O and Mg are, in general, underabundant and Si is always overabundant compared to the Sun. Remarkably, we found that the abundance of He increases during the main sequence lifetime from about 1\% to 10\% of the solar He abundance. We conclude that the observed increase is either the result of significant mixing in the stellar envelope of He that is not related to diffusion, or that there is accretion of He-rich gas from the interstellar medium.

As the climate in magnetic Bp stars changes with stellar age, important and systematic changes in surface chemistry result. These changes are due to both the evolving current climate of the outer layers of the star, and to the previous history of diffusion and competing effects. The systematic changes we have discovered present a major challenge to theoretical modelling. Efforts to reproduce theoretically the observed evolutionary changes should lead to greatly improved understanding of the physics at work in the envelopes of
magnetic Bp stars. Specifically, more detailed calculations on the equilibrium abundances for elements are necessary to fully interpret the results we present. Further, efforts to follow the evolution of a 3.5~M$_{\odot}$ star with time from a series of equilibrium abundance models will help differentiate between competing effects that influence the amount of radiative support for atoms in the stellar atmosphere. These competing effects include, for example, the decrease in gravity and the probable decrease in radiative acceleration with decreasing \te.

Future work will increase the number of stars in the current mass bin to enhance the current sample as well as increase the number of mass bins studied to include more massive stars (4-5~M$_{\odot}$) and less massive stars (2-3~M$_{\odot}$) to see if similar trends are observed.
\begin{acknowledgements}
JDB and JDL are grateful for support by the Natural Sciences and Engineering Research Council of Canada. JDL is pleased to acknowledge a Leverhulme Visiting Professorship at the Armagh Observatory. JDB thanks Dr. Tanya Ryabchikova for generously sharing her improved data to help create Figure 4.
\end{acknowledgements}

\bibliographystyle{aa}
\bibliography{abund-evol}

\end{document}